\documentclass[sigconf]{acmart}
\acmConference[ASE '22]{The 37th IEEE/ACM International Conference on Automated Software Engineering}{10 - 14 October, 2022}{Ann Arbor, MI, USA}

\AtBeginDocument{%
  \providecommand\BibTeX{{%
    \normalfont B\kern-0.5em{\scshape i\kern-0.25em b}\kern-0.8em\TeX}}}

\copyrightyear{2022}
\acmYear{2022}
\setcopyright{acmcopyright}\acmConference[ASE '22]{37th IEEE/ACM International Conference on Automated Software Engineering}{October 10--14, 2022}{Rochester, MI, USA}
\acmBooktitle{37th IEEE/ACM International Conference on Automated Software Engineering (ASE '22), October 10--14, 2022, Rochester, MI, USA}
\acmPrice{15.00}
\acmDOI{10.1145/3551349.3556948}
\acmISBN{978-1-4503-9475-8/22/10}

\usepackage{enumitem}
\usepackage{colortbl}
\usepackage{multirow}
\usepackage{subfigure} 
\usepackage{graphicx}
\usepackage{amsmath}
\usepackage{indentfirst}
\usepackage{alphalph}

\usepackage{booktabs} 
\newsavebox\CBox
\def\textBF#1{\sbox\CBox{#1}\resizebox{\wd\CBox}{\ht\CBox}{\textbf{#1}}}

\copyrightyear{2022}
\acmYear{2022}
\setcopyright{acmcopyright}\acmConference[ASE '22]{37th IEEE/ACM International Conference on Automated Software Engineering}{October 10--14, 2022}{Rochester, MI, USA}
\acmBooktitle{37th IEEE/ACM International Conference on Automated Software Engineering (ASE '22), October 10--14, 2022, Rochester, MI, USA}
\acmPrice{15.00}
\acmDOI{10.1145/3551349.3556948}
\acmISBN{978-1-4503-9475-8/22/10}

\begin{document}

\title{Using Consensual Biterms from Text Structures of Requirements and Code to Improve IR-Based Traceability Recovery} 

\settopmatter{authorsperrow=3}

\author{Hui Gao}
\email{ghalexcs@gmail.com}
\affiliation{%
  \institution{State Key Lab of Novel Software Technology, Nanjing University}
  \city{Nanjing}
  \country{China}
}

\author{Hongyu Kuang}
\email{khy@nju.edu.cn}
\affiliation{%
  \institution{State Key Lab of Novel Software Technology, Nanjing University}
  \city{Nanjing}
  \country{China}
}

\author{Kexin Sun}
\email{mf20320130@smail.nju.edu.cn}
\affiliation{%
	\institution{State Key Lab of Novel Software Technology, Nanjing University}
	\city{Nanjing}
	\country{China}
}

\author{Xiaoxing Ma}
\email{xxm@nju.edu.cn}
\affiliation{%
  \institution{State Key Lab of Novel Software Technology, Nanjing University}
  \city{Nanjing}
  \country{China}
}

\author{Alexander Egyed}
\email{alexander.egyed@jku.at}
\affiliation{%
  \institution{Institute for Software Systems Engineering, Johannes Kepler University}
  \city{Linz}
  \country{Austria}
}

\author{Patrick Mäder}
\email{patrick.maeder@tu-ilmenau.de}
\affiliation{%
  \institution{Fakultät für Informatik und Automatisierung, Technische Universität Ilmenau}
  \city{Ilmenau}
  \country{ Germany}
}

\author{Guoping Rong}
\email{ronggp@nju.edu.cn}
\affiliation{%
	\institution{State Key Lab of Novel Software Technology, Nanjing University}
	\city{Nanjing}
	\country{China}
}

\author{Dong Shao}
\email{dongshao@nju.edu.cn}
\affiliation{%
	\institution{State Key Lab of Novel Software Technology, Nanjing University}
	\city{Nanjing}
	\country{China}
}

\author{He Zhang}
\email{hezhang@nju.edu.cn}
\affiliation{%
	\institution{State Key Lab of Novel Software Technology, Nanjing University}
	\city{Nanjing}
	\country{China}
}

\begin{abstract}
  Traceability approves trace links among software artifacts based on whether two artifacts are related by system functionalities.
  The traces are valuable for software development, but are difficult to obtain manually.
  To cope with the costly and fallible manual recovery, automated approaches are proposed to recover traces through textual similarities among software artifacts, such as those based on Information Retrieval (IR).
  However, the low quality \& quantity of artifact texts negatively impact the calculated IR values, thus greatly hindering the performance of IR-based approaches.
  In this study, we propose to extract co-occurred word pairs from the text structures of both requirements and code (i.e., consensual biterms) to improve IR-based traceability recovery.
  We first collect a set of biterms based on the part-of-speech of requirement texts, and then filter them through the code texts.
  We then use these consensual biterms to both enrich the input corpus for IR techniques and enhance the calculations of IR values.
  A nine-system-based evaluation shows that in general, when solely used to enhance IR techniques, our approach can outperform pure IR-based approaches and another baseline by 21.9\% \& 21.8\% in AP, and 9.3\% \& 7.2\% in MAP, respectively. Moreover, when used to collaborate with another enhancing strategy from different perspectives, it can outperform this baseline by 5.9\% in AP and 4.8\% in MAP.
\end{abstract}




\maketitle


\section{Introduction and Motivation}
Software traceability is “the ability to interrelate any uniquely identifiable software engineering artifact to any other, maintain required links over time, and use the resulting network to answer questions of both the software product and its development process”\footnote{ http://www.CoEST.org}. 
For example, these \textit{traces} can reveal how stakeholders' expectations of system functionalities (i.e., requirements \cite{DBLP:journals/software/Cleland-Huang13}) are actually implemented and executed during the running of the system (i.e., code \cite{Binkley/SourceCodeAnalysis}).
Recent work has shown that, when correctly recovered, requirements-to-code traces can help developers to perform software maintenance tasks faster and more correctly\cite{DBLP:conf/icsm/MaderE12}, and the software quality is also highly relevant to the completeness of these traces\cite{DBLP:journals/tse/RempelM17}. 
In practice, traceability is not only mandated in certain regulations to demonstrate that a system is safely running \cite{DBLP:journals/infsof/NejatiSFBC12, DBLP:journals/software/MaderJZC13}, but also increasingly used to help ensure the security of a system \cite{DBLP:conf/icse/NhlabatsiYZTKBK15, DBLP:conf/icse/MoranPBMPSJ20}.
Unfortunately, the manual recovery of traceability is labor-intensive and error-prone \citep{DBLP:journals/tse/RameshJ01, DBLP:conf/re/EgyedGG10} due to the semantic gap between artifacts of different concept levels \cite{DBLP:conf/icse/BiggerstaffMW93} (such as requirements and code), and a large number of traces to recover.

To both improve the overall accuracy of traceability and reduce manual efforts, a growing body of automated approaches is proposed to recover or maintain traces based on mainly information retrieval (IR) and machine learning (ML) techniques. 
To bridge the semantic gap between different artifacts, these approaches typically use the calculated textual similarities of artifacts to represent the likelihood of the actual traces. 
Particularly, when recovering traces from scratch, i.e., no prior identified traces are available to form the training set of ML techniques, the semi-automated, IR-based approaches become the mainstream in research \cite{DBLP:conf/icse/Cleland-HuangGHMZ14}. 
Instead of verifying all possible traces between any two artifacts, users of these approaches can verify candidate trace links along the automatically generated candidate lists sorted by textual similarities (in descending order) that are calculated through IR models, such as Vector Space Model (VSM) \citep{DBLP:journals/tse/AntoniolCCLM02}, Latent Semantic Indexing (LSI) \citep{DBLP:conf/icse/MarcusM03}, and the probabilistic Jensen and Shannon model (JS) \citep{DBLP:conf/iwpc/AbadiNS08}.

\begin{figure*}[tb]
	\includegraphics[width=\textwidth]{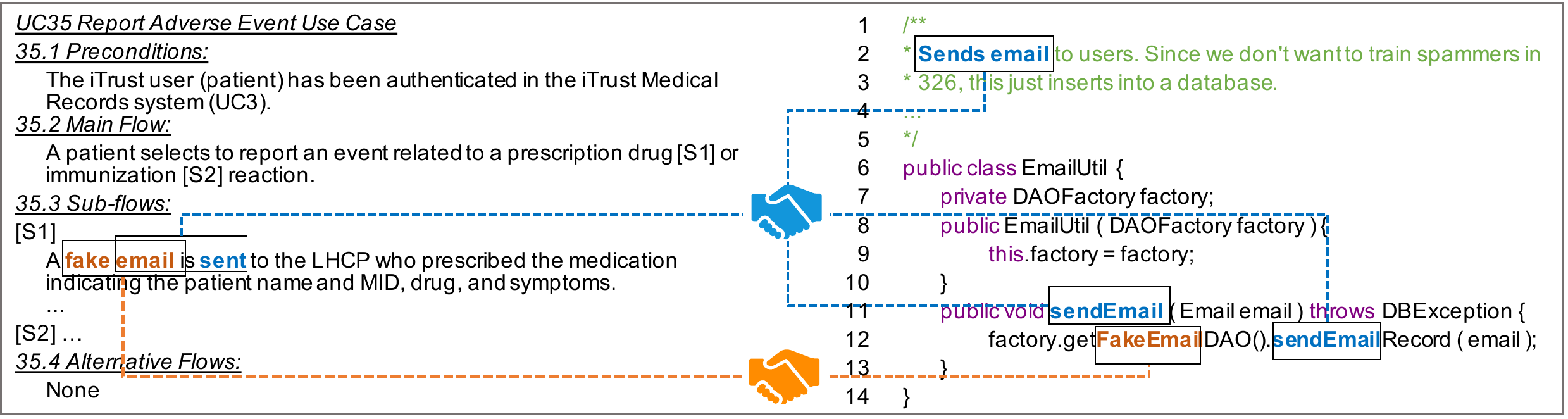}
	\caption{Motivating example adapted from the iTrust system}
	\label{fig1.1}
\end{figure*}

Unfortunately, different software artifacts (such as requirements and code) often use different terms to denote the same concept because they are at different concept levels \cite{DBLP:conf/icse/BiggerstaffMW93}. 
This situation, which is known as the \textit{vocabulary mismatch problem}, greatly hinders the performance of IR techniques on traceability recovery. 
Naturally, researchers proposed to use advanced lexical analyses to either enhance the similarity calculation through IR models \citep{DBLP:conf/icsm/GethersOPL11, Panichella6606598}, or enrich the artifact texts as the input of IR models \citep{DBLP:conf/re/Cleland-HuangSDZ05, DBLP:conf/iwpc/DiazBMOTL13}. 
However, the improvement brought by these approaches is limited to the unsatisfying quantity and quality of artifact texts, especially for real-world software projects. 
Therefore, researchers further proposed enhancing strategies that are built upon other unique perspectives of IR-based traceability recovery, such as incorporating with execution tracing \citep{DBLP:journals/tse/PoshyvanykGMAR07, DBLP:journals/ese/DitRP13} by running the systems, combining with the analyses on code structural information (e.g., call and data dependencies between code elements) \citep{DBLP:conf/icse/McMillanPR09, DBLP:conf/wcre/KuangNHRLEM17}, and utilizing user's verification on candidate links \citep{DBLP:journals/tse/HayesDS06, DBLP:conf/icsm/LuciaOS06, DBLP:conf/csmr/PanichellaMMPOPL13, DBLP:conf/icse/PanichellaLZ15, DBLP:journals/ese/GaoKMHLME22} due to the semi-automated nature of IR-based traceability recovery. 
But yet the performance of these enhancing strategies are still highly relevant to the quality of initially calculated IR similarities.

Meanwhile, a different body of work aims to reduce the “noise” from the texts of different artifact texts by focusing on the writing of software documents and code. 
De Lucia et al. \citep{DBLP:conf/iwpc/LuciaPOPP11} proposed a smoothing filter to remove the duplicate information in the same kind of artifacts (e.g., requirements or code) that does not carry the semantics of the artifact. 
Furthermore, Capobianco et al. \citep{DBLP:journals/smr/CapobiancoLOPP13} proposed that software artifacts are typically written by a “technical language” because the users of this language, i.e., the developers from the same software project, work in a particular area and have a common interest \citep{bookKeenan, bookJurafsky}. 
In this case, the nouns contained in the artifact texts are more likely to carry the semantics of the artifact and thus are more valuable to improve IR-based traceability recovery. Recently, Ali et al. \citep{articleAli2018} reported that not only nouns, but also the other major Part-of-Speech (POS) categories (including verbs, adjectives, adverbs, and pronouns) are important for IR-based traceability recovery. 
They further prune generated IR candidate links by finding whether the requirement and code from each given candidate link share at least one verb, to improve IR-based traceability recovery. 
Although these approaches alleviate the vocabulary mismatch problem by reducing the “noise” in artifact texts, they do not enhance, or even weaken the underlying artifact semantics (e.g., only consider nouns for IR techniques), thus preventing them from bringing in more improvement.

Unlike these approaches, in this paper, we propose to use \textit{consensual biterms} that are extracted from both the requirements and the code, to first enrich the artifact texts as the input for IR-based traceability recovery, and then use these extracted biterms to further improve the ranking of generated IR candidate lists. 
In particular, a \textit{biterm} is an unordered term-pair co-occurring over a term sequence within a text. 
It is first proposed to address the sparse data problem for retrieving documents through IR techniques \citep{10.1145/564376.564476BTIR}, and then used in topic modeling over short texts \citep{6778764BTM} to address the same problem. 
Specifically, we first extract an initial set of biterms from the requirement texts based on the grammatical relationships between two terms in each sentence by conducting the Stanford typed dependencies parsing \citep{DBLP:conf/acl/ManningSBFBM14}. 
We then extract another set of biterms from code identifiers and comments by using a sliding window. 
Finally, we keep the biterms from the intersection of these two sets and name them as \textbf{consensual biterms} because we argue that these biterms indicate at least part of the same system functionalities that are characterized in different software artifacts at different concept level, like requirements and code elements reach a “consensus” over these biterms. 
We then argue that extracting consensual biterms from common software projects is viable because as we discussed, requirements are written through a technical language that is though less formal than programming language, but still more rigorous than common natural language, and code is implemented through a well-defined programming language and the naming of its identifiers usually follows suggested conventions. 
For example, the Java Language Specification \footnote{Java language and virtual machine Spec. : https://docs.oracle.com/javase/specs/} recommends that class names should be nouns or noun phrases, and method names should be verbs or verb phrases (C Language also has the similar naming conventions \footnote{C Coding Standard:  https://users.ece.cmu.edu/\textasciitilde eno/coding/CCodingStandard.html}). 
Therefore, we propose that \textit{the consensual biterms extracted from the texts of both requirements and code can help to bridge the semantic gap between requirements and code, and thus they are very important to improve IR-based traceability recovery.} 
We will use the following example (adapted from the widely researched iTrust system \footnote{ https://agile.csc.ncsu.edu/iTrust/wiki/doku.php}) for further demonstration.

iTrust is a medicare system and one of its requirements (UC35) is to report adverse events on prescription drugs or immunization reactions through emails (though the email is never sent because iTrust is designed for lecturing testing courses only). 
Accordingly, the class \texttt{EmailUtil} is implemented to send those emails and thus it is traced to UC35. 
However, as depicted in Figure \ref{fig1.1}, UC35 only has one sub-flow S1 to describe the use of email. 
Thus, the textual similarity between UC35 and \texttt{EmailUtil} is inevitably small when calculated through IR models. 
Furthermore, the term “send” and its past participle “sent” are often filtered by the stop word list of IR-based approaches for traceability recovery because they appear too frequently in typical software artifacts and is considered as part of the “noise” in artifact texts. 
Unfortunately, without this term, the system functionality “send email” is also omitted by IR models as well, making the rank of \texttt{EmailUtil} very low in the IR candidate list for UC35. 
In contrary, we use the Stanford typed dependencies parsing to extract two biterms from the sub-flow S1, i.e., (send, email) and (fake, email), based on two grammatical relationships: \textit{noun subject} and \textit{adjectival modifier}, respectively. 
These two biterms can also be extracted from the method identifiers and the class comment of \texttt{EmailUtil}, and they become the consensual biterms. 
We argue that these two consensual biterms can narrow down the semantic gap between UC35 and \texttt{EmailUtil} by carrying the shared system functionality: “send fake email to report adverse event”, thus helping to improve IR-based traceability recovery.

Based on the consensual biterm, we further propose an enhancing strategy for IR-based, requirements-to-code traceability recovery. 
We first extracts consensual biterms from the texts of requirements and code, and then use them to enrich the corpus with other extracted terms for the following similarity calculations through IR models. 
Because consensual biterms carry the shared system functionalities between requirements and code, we further use them to adjust the generated IR value for a give candidate link by counting the occurrence of a consensual biterm globally (i.e., between a requirement and a code file) and locally (i.e., among different sections of a requirement, e.g., “title” and “sub-flow” for use cases, while “summary” and “description” for issues). 
Eventually, all IR candidate lists are re-ranked based on both the enriched corpus with consensual biterms and the adjustments of IR candidate links based on the occurrences of consensual biterms in code and the text structures of requirements. 
The evaluation of our proposed approach is based on nine real-world systems, and also involves the three mainstream IR models (i.e., VSM, LSI, and JS). 
Our evaluation first showed that our approach can statistically outperform the pure IR-based approach and the state-of-the-art IR-based approach that utilizes the POS tags for enhancement \citep{articleAli2018}. 
Our evaluation then showed that our approach, which is proposed by analyzing the writing of artifact texts, can also collaborate with another enhancing strategy \citep{DBLP:journals/ese/GaoKMHLME22} based on two different perspectives (i.e., combining user feedback with code dependency analysis), to further improve IR-based traceability recovery.

The contribution of this paper is extracting consensual biterms from the texts of requirements and code to both enrich the input corpora for IR models and improve the rankings of IR candidate lists. 
This work mainly targets functional requirements and focuses on tracing the requirements to code classes particularly. We name our approach as \textbf{TAROT} (\textbf{T}race\textbf{A}bility \textbf{R}ecovery by c\textbf{O}sensual bi\textbf{T}erms). 
This work contains two novel features: (1) we extract consensual biterms from requirements and code based on grammatical relationships (by using the Stanford CoreNLP toolkit) and naming conventions, respectively; (2) we improve the overall accuracy of IR-based traceability recovery through extracted consensual biterms to both enrich the input corpora for IR techniques and re-rank the IR candidate lists.
 

\section{Background and Related Work}

This section discusses the background and related work on typical approaches for traceability recovery, enhancing strategies for IR-based approaches, and the use of biterms in lexical analyses.

\textbf{IR-based traceability recovery:} Information-Retrieval (IR) -based approaches are perhaps now the most representative ones in automated traceability recovery \citep{DBLP:conf/icse/Cleland-HuangGHMZ14}. These approaches use IR techniques to compute the textual similarities between two different software artifacts (e.g., requirements and code), and suggest the IR values as the probability of whether a pair of source-target artifacts is a relevant trace. IR techniques compute the textual similarity based on the occurrence of terms in the text of artifacts.
If two artifacts share a larger number of terms, they will be more textually similar, and thus are more likely to be linked by a relevant trace.
In general, typical IR-based approaches use three steps to compute textual similarities \citep{DBLP:books/daglib/p/LuciaMOP12}: (1) they build a corpus with terms extracted from different software artifacts (the terms from code are usually extracted from identifiers and comments only) after pre-processing such as stemming and stop words removal; (2) these approaches use the term-by-document matrix to represent the corpus where each artifact are organized as a document in the matrix where its cell values are usually weighted by the tf-idf weighting scheme \citep{Baezayates2004Modern} to distinguish the importance for each term in the document based on its the occurring frequency; (3) these approaches compute the similarity among artifacts (represented as entries in the term-by-document matrix) through different IR models. 
The Vector Space Model (VSM) \citep{DBLP:journals/tse/AntoniolCCLM02} treats each document as a vector of terms. 
In this model, the similarity of two artifacts is typically calculated as the cosine of the angle between the two vectors in the space of terms occurred in all artifacts. 
The Latent Semantic Indexing (LSI) \citep{DBLP:conf/icse/MarcusM03} is a VSM that applies Singular Value Decomposition (SVD) \citep{Baezayates2004Modern} to the term-by-document matrix to construct an LSI subspace. 
The size of the subspace is a manually tuned threshold called the k value. Because the LSI subspace captures the most significant factors through SVD, LSI is expected to implicitly handle the most frequently co-occurring terms and thus to alleviate the vocabulary mismatch problem for IR-based traceability recovery.
Meanwhile, the Jenson-Shannon model (JS) \citep{DBLP:conf/iwpc/AbadiNS08} treats each document as a probability distribution containing the probability of a term occurring in each document. 
JS calculates the similarity among documents through the Jenson-Shannon divergence, i.e., the “distance” between two given probability distributions. 
To evaluate whether TAROT is generally beneficial to IR-based approaches, our evaluation involves all three discussed IR models.

\textbf{ML-based traceability recovery:} A growing body of work proposes to take advantage of Machine Learning techniques for traceability recovery. 
These ML-based approaches can work well especially when part of the trace links have been previous identified, such as completing the missing links between issues and code commits in software repository \citep{DBLP:conf/iwpc/LeVLP15, DBLP:conf/se/0002RGCM19, DBLP:journals/infsof/SunWY17}, or maintaining recovered traces through machine learning classification \citep{DBLP:conf/icsm/MillsEH18}. 
It is worth-while noting that due to the data imbalance problem between known traces and no-traces, re-balancing steps (e.g., under-sampling no-traces or oversampling traces) are necessary for these ML-based approaches. 
These approaches often use calculated IR values as the features of their classification, while our approach is expected to improve the quality of IR values, thus being likely to improve these approaches as well. 
Researchers also proposed deep-learning-based approaches, such as the one proposed by Guo et al. \citep{DBLP:conf/icse/0004CC17} that uses the RNN network to generate links between subsystem requirements and design definitions, and T-BERT \citep{DBLP:conf/icse/LinLZ0C21} that uses different BERT architectures to recover missing links between issues and commits. 
The deep learning (DL) network can capture the implicit connections from term sequences in artifact texts, while our approach extracts biterms explicitly from term sequences in software texts, and code identifiers and comments. 
However, the DL-based approaches still requires correctly-labeled training and development sets to work. 
To address this issue, Mills et al. \citep{DBLP:conf/icsm/MillsEBKCH19} introduce active learning to substantially reduce the amount of required training data for classification approaches while maintaining similar performance. 
Moran et al. \citep{DBLP:conf/icse/MoranPBMPSJ20} use a tailored hierarchical Bayesian Network to infer candidate traces through transitive relationships among different groups of software artifacts, and their approach Comet only needs a small amount of user feedback for better inference. 
Although we extract consensual biterms from the texts of both requirements and code, the extraction itself does not need any trace links to be previously identified. 
Thus, our approach, which is an enhancement for IR-based approaches, can still work when the user has to recover trace links from scratch.

\textbf{Enhancement for IR-based approaches:} To address the vocabulary mismatch problem that greatly hinders the performance of IR-based traceability recovery, researchers have proposed many enhancing strategies from different perspectives, such as introducing enhanced lexical analyses on the text of artifacts \citep{DBLP:conf/re/Cleland-HuangSDZ05, DBLP:conf/icsm/GethersOPL11,DBLP:conf/iwpc/LuciaPOPP11}, incorporating with execution tracing \citep{DBLP:journals/tse/PoshyvanykGMAR07, DBLP:journals/ese/DitRP13}, or combining with the analyses on different kinds of code dependencies \citep{DBLP:conf/icse/McMillanPR09,DBLP:conf/wcre/KuangNHRLEM17}. 
Meanwhile, because the semi-automatic nature of IR-based traceability recovery requires user verifications on the generated candidate links, a different body of work \citep{DBLP:journals/tse/HayesDS06,DBLP:conf/icsm/LuciaOS06, DBLP:conf/icse/PanichellaLZ15} uses user feedback (candidate links verified as either relevant links or false positives) to adjust the calculation of IR values. 
Panichella et al. \citep{DBLP:conf/csmr/PanichellaMMPOPL13} further combined the user feedback with code dependency analysis, and Gao et al. \citep{DBLP:journals/ese/GaoKMHLME22} proposed CLUSTER' that uses the closeness analysis on code dependencies \citep{DBLP:conf/wcre/KuangNHRLEM17} to improve IR-based approaches by propagating only a small amount of user feedback. 
Although the discussed approaches can improve the performance of IR-based traceability recovery, their provided benefits are still highly relevant to the quantity and quality of artifact texts, or the initially calculated IR values. 
Researchers tried to improve the quality of artifact texts by reducing the “noise” (discussed in Section 1), but the remaining texts are not enough to further improve IR-based approaches. 
In contrary, we use extracted consensual biterms to first enrich the corpus for IR techniques, and then adjust the ranking of candidate lists without considering code dependencies or user feedback. 
We then combined TAROT with CLUSTER' in our evaluation, and the experiment results showed that TAROT can collaborate with other enhancing strategies for further improvement.

\begin{figure}[t] 
	\centering
	\includegraphics[width=\linewidth]{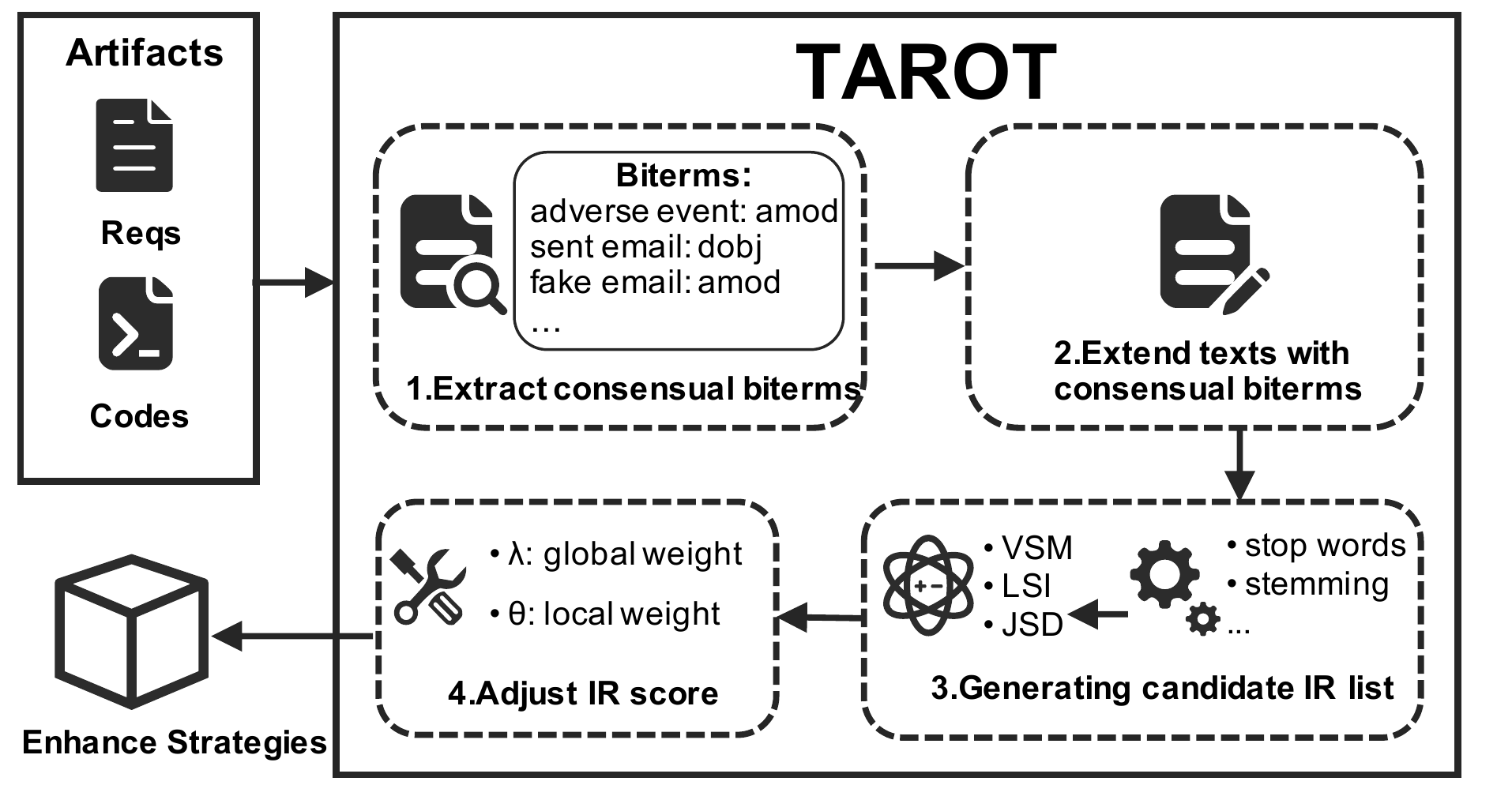}
	
	\caption{Overview of the TAROT framework.}
	\label{fig3.1}
\end{figure}

\textbf{Using biterms in lexical analyses:} Biterms are first proposed for document retrieval \citep{10.1145/564376.564476BTIR} to address the data sparsity problem. 
For the same reason, Cheng et al. \citep{6778764BTM} use biterms to better establish topic models over short texts. 
Although the use of biterms in software engineering (SE) research is not many, researchers seem to agree that biterms are useful in analyzing short SE texts. 
For example, Hadi et al. \citep{9240699} proposed an adaptive online biterm topic model to analyze version sensitive short texts. 
Instead of directly using the biterm topic modeling, we use biterms as the vital role of our approach to capture the same system functionalities that are described in different artifacts at different concept levels.

\section{Proposed Approach}
In this section, we introduce the proposed four-step approach TAROT. 
Figure \ref{fig3.1} illustrates the overview of TAROT.
First, we extract consensual biterms from both requirements and code. 
Second, we enrich artifact texts with consensual biterms (Section \ref{sec3.2}).
Third, we generate candidate trace lists  (Section \ref{sec3.3}).
Fourth, we further adjust IR values with global and local weight, i.e., $\lambda$ and $\theta$  (Section \ref{sec3.4}). 
The updated candidate trace lists can also be the input for other enhancing strategies, e.g., our baseline CLUSTER' \cite{DBLP:journals/ese/GaoKMHLME22}.

\subsection{Extracting Consensual Biterms} \label{sec3.1}
\subsubsection{Extracting Candidate Biterms from Requirements} \label{sec3.1.1}
\begin{figure}[t]
	\includegraphics[width=180pt]{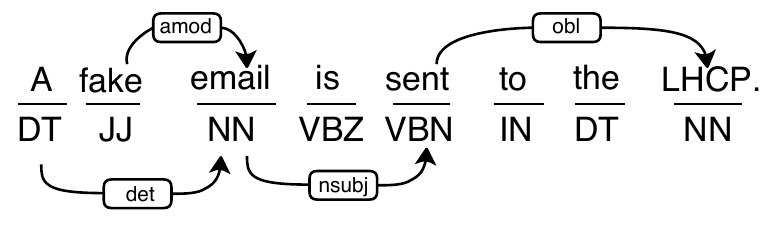}
	\caption{Stanford typed dependencies parsing in a sentence}
	\label{fig3.2}
\end{figure}

Two typical kinds of structured texts are often used to describe requirements in recent software development, i.e., use cases and repository issues.
The use case usually consists of five parts, including title, precondition, main-flow, sub-flow, and alternative-flow, while
the issue only has two parts: summary and description.
We first split each requirement into several parts  according to its text structure.
Because requirements are written with natural language, we leverage Standford CoreNLP \citep{DBLP:conf/acl/ManningSBFBM14} to  process each splitted part of text as follows:
(1) we split text into independent sentences; 
(2) we tag each term's Part-of-speech (POS, e.g, nouns, verbs, adjectives, adverbs, and pronouns) in the sentence;
(3) we conduct a Stanford typed dependencies parsing for each sentence that we can get the grammatical relationship between two terms in the sentence.
For example, given a sentence as shown in Figure \ref{fig3.2}, each term's POS tag is labeled under itself.
Stanford CoreNLP provides many types of tag, e.g., DT  (determiner), JJ (adjective), NN (noun), VBZ (verb, third person singular present), VBN (verb, past tense), and IN (preposition or subordinating conjunction), etc.
The grammatical relationship between two different terms is connected with arrowed lines, and relationships are labeled on the lines.
For example, “fake” and “email” have an \textit{amod} (adjectival modifier) relationship,
“email” and “sent” have a \textit{nsubj} (noun subject) relationship, and “sent” and “LHCP” (i.e., Licensed Health Care Professional) have a \textit{obl} (oblique nominal) realtionship.
Each two terms with grammatical relationship is recognized as a biterm (i.e., “\textbf{aEmail}”, “\textbf{fakeEmail}”, “\textbf{emailSent}”, and “\textbf{sentLHCP}”).
However, biterms are likely to carry different information values that depend on composed terms' POSs.
On the one hand, previous studies \citep{DBLP:conf/clef/KarimpourGPMAAO08, DBLP:journals/re/MahmoudN15} have shown that verbs, nouns, and adjectives play important roles in expressing semantics.
On the other hand, names of code entities (e.g., class name, method name, and variable name and type, etc.) are also generally composed of terms with the three POSs \cite{DBLP:conf/iwpc/PerumaHCAMN21, DBLP:journals/jss/NewmanADPKM020}.
Therefore, each term of a biterm should only be verb, noun, or adjective.
Otherwise,  the biterm will be ignored (e.g., “\textbf{aEmail}”).

\subsubsection{Extracting Candidate Biterms from Code} \label{sec3.1.2}
We extract biterms from both code identifier names (i.e., class name, method name, invoked method name, field name and its type, and parameter name and its type) and comments (i.e., comments for class or method). 
Unfortunately, existing POS taggers cannot achieved an accuracy of 100\% in tagging  code identifiers \cite{DBLP:journals/jss/NewmanADPKM020, DBLP:conf/iwpc/GuptaMPV13,DBLP:journals/infsof/AliCHH19}.
This is because code identifiers are not written in a natural language like requirements, so they do not contain the contexts of proper English sentences and grammatical structures.
Nevertheless, to help developers better comprehend code, identifier names are usually composed of two or more terms (usually no more than five terms \cite{DBLP:journals/jss/NewmanADPKM020,DBLP:conf/iwpc/ButlerWY15}) by following typical naming conventions, such as the \texttt{CamelCase} or the \texttt{snake\_case}. 
Furthermore, identifier names tend to follow certain POS-based patterns among its splitted terms in general~\cite{DBLP:journals/jss/NewmanADPKM020,DBLP:conf/msr/BinkleyHL11,DBLP:conf/iwpc/PerumaHCAMN21}, such as \textbf{VB + JJ + NN} (e.g., \texttt{getFakeEmail}) and \textbf{VB + NN} (e.g., \texttt{sendEmail}) in \texttt{EmailUtil} shown in Figure \ref{fig1.1}.
Therefore, we extract candidate biterms from identifier names by  combining any two splitted terms sequentially.
Take the identifier name \texttt{getFakeEmail} for example, we will extract three biterms including “\textbf{getFake}”, “\textbf{getEmail}”, and “\textbf{fakeEmail}”.
Although some unmeaning biterms may be extracted, i.e., “\textbf{getFake}”, we consider them still tolerable because they are not many in one given identifier name.
Furthermore, they are likely to be discarded when we further extract consensual biterms between requirements and code (discussed in Section \ref{3.1.3}).
Meanwhile for code comments, we conduct the same candidate biterm extractions on them as those on requirements because they are both written in natural languages.

\subsubsection{Extracting Consensual Biterms} \label{3.1.3}
We now use a quite straightforward crosscheck to extract the final consensual biterms from the two sets of candidate biterms from requirements and code, respectively.
First, to address the unmeaning biterm problem from code candidate biterms (mentioned in Section \ref{sec3.1.2}), we only retain biterms that also appear in the set of requirement candidate biterms.
The reason why we favor requirement candidate biterms is that we extract  these biterms based on both the grammatical dependencies and the selected types of POSs (discussed in Section \ref{sec3.1.1}), so that the quality of requirement candidate biterms is likely to be much higher, let alone the typically better quality and quantity of requirement texts to guarantee the quality of Stanford CoreNLP output.
Then we filtered out biterms that only appear in the set of requirement candidate biterms, to get the final set of consensual biterms between requirement and code. It is worth-while noting that each term in a given candidate biterm is restored to its original format, and the sequence of two terms is also omitted, except the domain abbreviations such as "LHCP" (which will be treated as a special noun). For example in our iTrust Sample, the "\textbf{emailSent}" mentioned in Section~\ref{sec3.1.1} and the "\textbf{sendEmail}" mentioned in Section~\ref{sec3.1.2} will both be saved as \textbf{(send, email)}, and this biterm becomes a consensual biterm after the crosscheck.

\subsection{Enriching Texts with Consensual Biterms} \label{sec3.2}
\subsubsection{Enriching Requirements with Consensual Biterms} \label{sec3.2.1}
To validly enrich requirement texts with extracted consensual biterms, we need to not only just "add them in the text", but also assign reasonable occurrence numbers for each added biterm so that our approach can effectively index these "artificial terms" later in calculating IR values between requirements and code classes. Specifically, we first record numbers of the occurrences of each consensual biterm respectively in different parts of a requirement.
Then we accumulate the numbers in each part as the total number that the consensual biterms should be added into the requirement.

\subsubsection{Enriching Code with Consensual Biterms} \label{sec3.2.2}
Similarly but differently, we design different count strategies according to the kind of identifier name in which the consensual biterm occurred.
First, if a biterm appears in class names or method names, its count pluses two.
This is because the names of classes and methods are particularly important.
Høst et al. \citep{DBLP:conf/ecoop/HostO09} considered that “Methods are the smallest named units of aggregated
behavior in most conventional programming languages and hence the cornerstone of abstraction”.
In addition, a class name is a high-level abstraction and summarization of a class, and a method name is a direct description of functionality.
Then for a biterm appears in comments (i.e., comments of class and methods), its count pluses one for each occurrence.
For invoked method name, field type and name, and parameter type and name, we take a conservative strategy.
If the biterm only appears in these parts, its count will be assigned to one and not change no matter how many times it appears.
Otherwise, we skip this biterm.
Finally, we accumulate all counts of a biterm as the total number that should be added to each code class.

\subsection{Generating Candidate Trace Links} \label{sec3.3}
\textbf{Creating Corpus}. 
Each class in code is extracted into one document containing its comments and identifiers including class names, method names, and field names. 
Then we use the CamelCase convention to split identifiers into terms. 
For each requirement, we extract a document that includes its title and content (e.g., preconditions, main-flow, sub-flows, and alternative-flows for structured use cases or title and description for commit issues). 

 \textbf{Normalizing Corpus}. 
 We normalize the documents of requirements and classes by standard pre-processing techniques for IR including: (1) removing special tokens (e.g., punctuations); (2) converting all upper case letters into the lower case; (3) removing stop words; (4) performing the Porter stemming algorithm \citep{DBLP:journals/program/Porter80}.

\textbf{Indexing Corpus and Computing Textual Similarities}. 
We use tf-idf for corpus indexing (for both terms and biterms) and three mainstream IR models to compute textual similarity: Vector Space Model (VSM) \citep{DBLP:journals/tse/AntoniolCCLM02}, Latent Semantic Indexing (LSI) \citep{DBLP:conf/icse/MarcusM03}, and the probabilistic Jensen and Shannon (JS) model \citep{DBLP:conf/iwpc/AbadiNS08}.

\textbf{Generating Candidate Links}. We rank IR candidate lists in descending order based on the IR values of candidate links.

\subsection{Adjusting IR Values through the Biterms} \label{sec3.4}
In addition to enriching artifact texts with biterms, we further leverage consensual biterms' global (i.e, for both a requirement and a code file) and local (i.e., for done part of a requirement) weights to adjust IR values to improve the ranking of generated IR candidate lists.
Given a requirement \textit{req} and a code \textit{cls}, we denote the set of biterms in \textit{req} and \textit{cls} as \textit{BT$_{req}$} and \textit{BT$_{cls}$} respectively.
Then we can get the set of biterms that they shared denoted as 
\begin{equation}
	BT_{cons} = BT_{req} \cap BT_{cls}, \hspace{1em}req \in UC\cup ISSUE, cls \in SC
\end{equation}	
where \textit{UC}  and  \textit{ISSUE} represent two different types of requirements, i.e., use cases and commit issues, respectively.
\textit{SC} represents all source code.
Note that we adopt each biterm's \textit{inverse document frequency} (IDF) as its weight due to it can provide a good measure of biterms' uniqueness.
For example, if a biterm occurred in almost all artifacts that are considered as a common biterm, then its IDF will be assigned a small value.

\begin{table*}[t]
 \caption{Overview of the Nine Evaluated Systems}
 \label{tab4.1}
 \small
 	\renewcommand\arraystretch{0.8}
 \begin{tabular}{lccccccccc}
 	
 	\toprule
 	\textbf{\#}                    & \textbf{iTrust}         & \textbf{\begin{tabular}[c]{@{}c@{}}Gantt\\ Project\end{tabular}}        & \textbf{Maven}        & \textbf{Pig}      & \textbf{Infinispan}      & \textbf{Seam2}        & \textbf{Drools}        & \textbf{Derby}          & \textbf{Groovy}          \\
 	\midrule
 	
 	\textbf{Version}       & 13.0            & 2.0.9         & 3.5.2           & 0.17.0         & 9.2.0      & 7.5.0      & 14.1.0        & 2.3.0      & 2.5.0    \\
 
 	\textbf{Programming Language}  & Java          & Java         & Java        & Java        & Java       & Java    & Java      & Java     & Java        \\
 	
 	\textbf{KLoC}           & 43      & 45       & 101     & 365  & 521    & 449     & 1091      & 136          & 381           \\
 
 	\textbf{Executed classes}  & 137        & 124    & 82      & 289    & 319      & 150      & 248     & 611           & 100               \\

 	\textbf{Evaluated requirements}     & 34    & 16    & 36     & 87   & 232           & 189        & 183         & 390   &104   \\             
 	
 	\textbf{Relevant Traces in RTM}    & 255       & 315        & 151    & 547   & 1116          & 463       & 841        & 2315     & 180           \\
 	
 	\bottomrule                                   
\end{tabular}
\end{table*}

\subsubsection{Global Weight $\lambda$} \label{3.4.1}
For a given requirement \textit{req} and code \textit{cls}, we consider more biterms they shared, \textit{req} and \textit{cls} are more likely to be an actual trace link.
In addition to the size of $BT_{cons}$, we also take into  account the proportion of  $BT_{cons}$.
We calculate proportions of $BT_{cons}$ in both $BT_{req}$ and $BT_{cls}$ respectively.
Then, we combine these two proportions together and take half of the value as the global weight of $BT_{cons}$ denoted as $\lambda$:
\begin{equation}
	\lambda(req, cls)  =(\frac{\sum\nolimits_{k=1}^{|BT_{cons}|}IDF_{bt_k}}{\sum\nolimits_{i=1}^{|BT_{req}|}IDF_{bt_i}}  + \frac{\sum\nolimits_{k=1}^{|BT_{cons}|}IDF_{bt_k}}{\sum\nolimits_{j=1}^{|BT_{cls}|}IDF_{bt_j}} )\times \frac{1}{2} 
\end{equation}
where $|BT_{req}|$,  $|BT_{cls}|$, and  $|BT_{cons}|$ represent  the number of biterms in $req$, $cls$, and their shared biterms, respectively. 
$IDF_{bt_i}$, $IDF_{bt_j}$, and $IDF_{bt_k}$ are the IDFs of the $i_{th}$ , $j_{th}$ , and $k_{th}$ biterm in $BT_{req}$, $BT_{cls}$, and $BT_{cons}$, respectively.

\subsubsection{Local Weight $\theta$} \label{3.4.2}
We further consider that each part of a requirement can differently contribute to the described system functionalities. We mainly consider two kinds of requirement artifacts: use cases and issues in software repositories.
For use case, title and main-flow are summarized descriptions of the requirement. 
Therefore, they hold more information values and biterms in these two parts should be assigned higher weights, especially for title since it is more condensed. 
Sub-flows describe the main scenario involving detailed functionalities  step-by-step.
Alternative-flows describe alternate ways that the system behaves according to specific inputs.
Precondition explains the state that the system must be in for the use case to be able to start which is not involved in functional description, so we do not consider this part.
We then refer to previous study \citep{DBLP:conf/icsm/AliSGA12} and assign the following weights 0.4, 0.3, 0.2, and 0.1 to title, main-flow, sub-flow, and alternative-flow, respectively.
Similarly, we assign weights 0.6 and 0.4 to summary and description, respectively for repository issues.
Given a requirement $req$ and code $cls$, we compute the proportion of biterms they shared in all biterms that each part owns as follow:
\begin{equation}
	\omega(req\_part, cls)= \frac{\sum\nolimits_{k=1}^{|BT_{cons\_part}|}IDF_{bt_k}}{\sum\nolimits_{i=1}^{|BT_{req\_part}|}IDF_{bt_i}}
\end{equation}
where $req\_part$ is  one part of $req$, $|BT_{cons\_part}|$ is the number of shared biterms for $req_{part}$  and $cls$. I$DF_{bt_k}$, and $IDF_{bt_i}$ are the IDFs of the $k_{th}$ , $i_{th}$ in $BT_{req_part}$ and $BT_{cls}$ respectively. 
We then compute \textit{local weight} (denoted as $\theta$) by combining weights from different part of requirement texts as follows:
\begin{equation}
	\theta (req,cls)=\left\{
	\begin{aligned}
		&0.4 \times  \omega(title, cls) + 0.3 \times \omega(mf, cls) + \\
		&0.2 \times \omega(sf, cls) + 0.1 \times \omega(af, cls), req \in UC\\
		&\\
		&0.6 \times \omega(summ) +0.4 \times \omega(desc), \hspace{0.5em}req \in Issue
	\end{aligned}
	\right.
\end{equation}
where $title$ refers to "title", $mf$ refers to "main-flow", $sf$ refers to "sub-flow", and $af$ refers to "alternative-flow" in use cases, while $summ$ is "summary" and $desc$ is "description" in issues. 

Finally, we can leverage $\lambda$ and $\theta$ to adjust IR values if a requirement and code shared consensual biterms, i.e., $BT_{cons} \neq \emptyset $.
Otherwise, we will penalize the IR values through multiplying origin IR value by 0.9 as follows:
\begin{equation}
	IR_{new}=\left\{
	\begin{aligned}
		&IR_{initial} \times (1+\lambda+\theta), &&BT_{cons} \neq \emptyset \\
		&IR_{initial} \times 0.9, &&BT_{cons} =\emptyset 
	\end{aligned}
	\right.
\end{equation}
where $IR_{initial}$ is the initial IR value and $IR_{new}$ is the updated one.


\section{Experiment Setup}
We now introduce our experimental setup to evaluate our approach. 
Section \ref{sec4.1} introduces nine evaluated systems.
Section \ref{sec4.2} defines metrics for evaluating the performance of our approach and baseline approaches. 
Section \ref{sec4.3} discusses our research questions and the design of experiments.

\subsection{Evaluated Systems}  \label{sec4.1}
Our evaluation is based on nine real-world software systems: iTrust, GanttProject\footnote{https://github.com/bardsoftware/ganttproject}, Maven\footnote{https://github.com/apache/maven}, Pig\footnote{https://github.com/apache/pig}, Infinispan\footnote{https://github.com/infinispan/infinispan}, Drools\footnote{https://github.com/kiegroup/drools}, Derby\footnote{https://github.com/apache/derby}, Seam2\footnote{http://www.seamframework.org/Seam2.html}, and Groovy\footnote{https://github.com/apache/groovy}. 
Table \ref{tab4.1} presents a summary of nine evaluated systems.
We chose these systems because of their availability of both requirements specifications with developer maintained use cases and their Requirements-to-code Trace Matrices (RTM). For GanttProject, high-quality class-level traces are gained by recruiting the original developers. The RTM of iTrust contains method-level traces maintained by original developers and is publicly available. However, the RTMs of the other eight systems are class-level. To keep our experiment consistent at the same granularity, we propagated the method-level traces of iTrust to class-level traces by aggregating all traces to methods of a class on the class-level.

Meanwhile, Maven, Pig, Infinispan, Drools, Derby, Seam2, and Groovy come from the dataset named IlmSeven \citep{DBLP:conf/re/RathRM17}. 
The dataset consists of seven open source projects where the traces between issues and changed classes for each project are elicited from its GitHub and the issue-tracking tool Jira. 
Specifically, Maven is the mainstream software project management tool.
Pig is a platform to analyze large datasets consisting of a high-level language for expressing data analysis programs along with the infrastructure for assessing these programs. 
Infinispan is a distributed in-memory key-value NoSQL data store software.
Drools is a business rule management system and reasoning engine for business policy and rules development, access, and change management.
Derby is a relational database management system that can be embedded in Java programs and used for online transaction processing. 
Seam2 is a lightweight framework for building web applications in Java.
Finally, Groovy is an object-oriented programming language for the Java platform.
All projects are implemented in the Java language (the Groovy project using both Java and Groovy itself). 
We used all seven projects and follow the same suggestions and heuristics proposed by Hui et al. \citep{DBLP:journals/ese/GaoKMHLME22} (also our baseline) to filter and merge each project's issues to avoid the likely noises and too fine-grained system functionalities when using the original issues as requirements.
Our dataset is available at: https://github.com/huiAlex/TAROT.

\subsection{Evaluation Metrics}  \label{sec4.2}
We first leveraged two well-known metrics (i.e., recall and precision) to evaluate the performance of our approach. Precision represents the proportion of correct links among retrieved trace links, and recall represents the proportion of retrieved correct links among all correct links. They are defined as follows:\\
\begin{equation}
	Precision = \frac{\left | relevant\cap retrieved\right |}{\left | retrieved\right |}
\end{equation}
\begin{equation}
	Recall = \frac{\left | relevant\cap retrieved\right |}{\left | relevant\right |}      
\end{equation}
where $relevant$ is the set of relevant links and $retrieved$ is the set of links retrieved by traceability recovery approaches.

A common way to evaluate the accuracy of IR techniques is to compare the precision values obtained at different recall levels, resulting in a set of precision-recall points displayed as curves. We then leveraged the following two metrics: AP (i.e., average precision) and MAP (i.e., mean average precision). 
AP and MAP are computed as:
\begin{equation}
	AP = \frac{\sum_{r=1}^{N}(Precision(r)\times isRelevant(r))}{\left | RelevantDocuments\right |}
\end{equation}
\begin{equation}
	MAP = \frac{\sum_{q=1}^{Q}AP(q)}{Q}
\end{equation}
where $r$ is the rank of the target artifact in an ordered list of links, $Precision(r)$ represents its precision value, $isRelevant()$ is a binary function assigned 1 if the link is relevant or 0 otherwise, $N$ is the number of all documents, $q$ is a single query, and $Q$ is the number of all queries. AP measures how well relevant documents of all queries (requirements) are ranked to the top of the retrieved links. Meanwhile, MAP uses the average of the AP scores of all queries to measure how well relevant documents for each query are ranked to the top of the retrieved links.

\subsection{Research Questions}  \label{sec4.3}
We aim to study whether the use of consensual biterms from text structures of requirements and code, i.e., enriching the input corpus and adjusting the calculated IR values, can help improve IR-based traceability recovery.
Furthermore, because TAROT is proposed based on text structures only, we also want to study whether it can collaborate with other enhancing strategies built upon different perspectives (e.g., code dependency analysis and the use of user feedback). Finally, we want to study how the corpus-enriching part and the value-adjusting part contribute  separately to TAROT. 
Thus, we proposed the following three research questions:

\vspace{5pt}

\textbf{RQ1:} \textit{Can TAROT help improve the performance of IR-based traceability recovery?}

\textbf{RQ2:} \textit{Can TAROT collaborate with other enhancing strategies built upon different perspectives for IR-based traceability recovery?}

\textbf{RQ3:} \textit{How many contributions do different parts of TAROT make individually?}

\vspace{5pt}
To study RQ1, we combine TAROT with the pure IR-based approach and named the combination as \textbf{IR-ONLY\_TAROT}.
We then compared the performance of IR-ONLY\_TAROT and two baseline approaches.
The first is the pure IR-based approach (\textbf{IR-ONLY}, without any enhancing strategies involved).
Another one is \textbf{ConPOS}, which is the state-of-the-art approach that uses major POS categories and applies constraints to prune candidate IR links as a filtering process \citep{articleAli2018}, while TAROT also uses POS taggers, but it chooses to first enrich the input corpora for IR techniques and then adjust the calculated IR values. We argue that TAROT have more potentials to better improving IR-based approaches because the quantity of either requirement texts or code texts is already quite small in practice. Correctly enriching those texts will be eventually more effective than further pruning them, and ConPOS is a really good baseline to evaluate our idea and approach.
We then involved three mainstream IR models in our evaluation, i.e., VSM, LSI, and JS (which are widely used and evaluated in traceability research~\cite{DBLP:conf/icse/Cleland-HuangGHMZ14}), to compare  IR-ONLY\_TAROT with the two baseline approaches.
Note that for the k value of LSI, we adopt a brute-force search strategy\cite{DBLP:conf/iwpc/KhatiwadaTM20} to determine the optimal value of k for each evaluated system.
For iTrust, GanttProject, Maven, Pig, Infinispan, Seam2, Drools, Derby, and Groovy, their calibrated k values are 85, 65, 70, 140, 170, 180, 300, 130, and 175, respectively. 
For new systems, we suggest using the same calibration process before fine-tuning them to optimize the performance.

In addition to the metrics proposed in Section \ref{sec4.2}, we also use a statistical significance test to check whether the performance of  IR-ONLY\_TAROT is significantly better than that of baseline approaches.
Through the significance test used in reference \cite{DBLP:journals/ese/AliSGA15, DBLP:conf/wcre/KuangNHRLEM17, DBLP:conf/iwpc/KuangG0M0ME19}, we use the F-measure at each recall point as the single dependent variable of our significant test. 
We use the F-measure because we want to know whether IR-ONLY\_TAROT improves both accuracy and recall. 
F-measured value is calculated as follows:

\begin{equation}
	F=\frac{2P\times R}{P+ R}
\end{equation}
where \textit{P} represents precision, \textit{R} represents recall, and \textit{F} is the harmonic mean of \textit{P} and \textit{R}. 
A higher F-measure means that both precision and recall are high and balanced. 
We then use the Wilcoxon rank sum test \citep{WilcoxonIndividual} to test the following null hypothesis:

\textit{$H_0$: There is no difference between the performance of TAROT and baseline approaches.}

We use $\alpha$ = 0.05 to accept or refute the null hypothesis $H_0$. 
We also use a non-parametric effect size measure for ordinal data, i.e., Cliff's $delta$ ($\delta$) \citep{MacbethCliff}, to quantify the amount of difference between  IR-ONLY\_TAROT and two baseline appproaches as follows:

\begin{equation}
	\delta=\left |  \frac{\#\left ( x_{1}> x_{2}\right ) - \#\left ( x_{1}<  x_{2}\right )}{n_{1}n_{2}}   \right |
\end{equation}
where $x_1$ and $x_2$ represent F-measures of TAROT and a chosen baseline approach, and $n_1$ and $n_2$ are the sizes of the sample groups. 
The effect size is considered 
negligible for $\delta < 0.15$, 
small for $0.15 \leq \delta < 0.33$, 
medium for $0.33 \leq \delta < 0.47$, 
large for $\delta > 0.47$.

\begin{table*}[tb]
	\caption{The number of computed AP, MAP, p-value, and Cliff’s $\delta$ evaluating each approach (evaluated systems iTrust, GanttProject, Maven, Pig, Infnispan, Seam2, Drools, Derby, and Groovy combined with IR models VSM, LSI, and JS)}
	\renewcommand\arraystretch{0.6}
	\resizebox{\textwidth}{!}{
	\label{tab5.1}
	\tiny
	\begin{tabular} {ll cccc cccc cccc}
		\toprule
		&         & \multicolumn{4}{c}{VSM} & \multicolumn{4}{c}{LSI} & \multicolumn{4}{c}{JS} \\ 
		\cmidrule(r){3-6} 	\cmidrule(r){7-10}	\cmidrule(r){11-14}
		&          &AP& MAP    & $p$-value 		& C'$\delta$       &AP& MAP    & $p$-value       		& C'$\delta$       &AP& MAP  & $p$-value    		& C'$\delta$    \\ 
		\midrule
		
		\multirow{6}*{iTrust}  
		& IR-ONLY		        	 &45.78& 58.43  & \textBF {0.05}& 0.10     				&46.01& 59.17 & 0.06 & 0.10             					&40.57& 56.01& \textBF{\textless 0.01} & 0.14   \\
		& ConPOS                    &46.89& 59.03 &0.13 & 0.08                      		 &46.60& 59.00 & 0.17& 0.07           		 			    	&39.95& 55.27& \textBF{0.02}&0.12 \\
		& IR-ONLY\_TAROT	  &\textBF{49.50}& \textBF{62.12}& -& -          		  &\textBF{49.18}& \textBF{61.50} & -& -                &\textBF{45.74}&\textBF{58.59}& -& -   \\
		\cmidrule{2-14}
		& CLUSTER'                 & 64.08& 68.51  & 0.54 &0.03       						 &64.55& 69.94  &  0.34  & 0.05          		      &61.75&66.53  &  \textBF{0.03}   & 0.11 \\
		& CLUSTER'\_TAROT   & \textBF{65.90}& \textBF{72.33}  & -& -          		&\textBF{67.02}& \textBF{72.95}  & -& -            &\textBF{64.20}& \textBF{69.09}  & -& - \\
		\midrule
		
		\multirow{6}*{GanttProject}    
		& IR-ONLY      				&43.17& 49.79& \textBF{\textless 0.01}& 0.14   		&43.89	& 51.72 & \textBF{\textless 0.01} & 0.15   	&36.50& 46.76& \textBF{\textless 0.01} & 0.22     \\
		& ConPOS                   &44.33& 51.41  & \textBF{\textless 0.01}& 0.17       &44.81& 53.15 & \textBF{\textless 0.01}& 0.18         &38.24& 50.04& \textBF{\textless 0.01}&0.24 \\
		& IR-ONLY\_TAROT  	 &\textBF{47.37}& \textBF{54.09} &-& -                   &\textBF{48.63}& \textBF{55.25} &-& -                    &\textBF{43.90}&\textBF{51.76} &-& -  \\
		\cmidrule{2-14}
		&  CLUSTER'   			  &62.49&71.90 &\textBF{\textless 0.01}  & 0.18       &66.28&\textBF{72.13}  &0.09  &  0.08                                 & 70.69& 74.10& 0.97  & 0.0  \\
		& CLUSTER'\_TAROT  &\textBF{73.18}& \textBF{75.38} &-& -         	        &\textBF{70.91}& 72.05&-& -           		   &\textBF{73.19}&\textBF{75.37}&-& -    \\
		\midrule
		
		\multirow{6}*{Maven} 
		& IR-ONLY  				    &22.27& 37.11& \textBF{\textless 0.01}& 0.24           &22.15&40.65 & \textBF{\textless 0.01} & 0.20              &24.08& 40.29 & \textBF{\textless 0.01} & 0.29        \\
		& ConPOS                   &22.34& 37.12  & \textBF{\textless 0.01}& 0.17         &21.23& 38.67 & \textBF{\textless 0.01}& 0.18            &22.56& 37.33& \textBF{\textless 0.01}&0.24 \\
		& IR-ONLY\_TAROT  	 &\textBF{26.77}& \textBF{43.65} & -& -         			  &\textBF{26.97}& \textBF{47.17} & -& -          			&\textBF{25.50}&\textBF{48.35}& -& -    \\
		\cmidrule{2-14}
		& CLUSTER'  			  &  42.53& 47.31 &  0.61&  0.03           					   & 41.92&  49.61  &  0.75&  0.02           				  & 42.00&  49.95 &  0.21&  0.08  \\	
		& CLUSTER'\_TAROT  &\textBF{45.17}&\textBF{53.91} & -& -              		 &\textBF{44.18}& \textBF{54.87} & -& -            		 &\textBF{47.55}& \textBF{55.75}& -& -  \\
		
		\midrule
		
		\multirow{6}*{Pig} 
		& IR-ONLY    	           &19.71& 36.37&\textBF{0.02} & 0.08           			&17.89& 36.62  & 0.12& 0.05          			 &14.64& 31.90 & \textBF{\textless 0.01} & 0.13     \\
		& ConPOS              	  &20.05&39.18 &\textBF{0.03}& 0.07           				&17.96& 37.67 & 0.13& 0.05            			&13.46& 32.94&  \textBF{\textless 0.01} &0.13 \\
		& IR-ONLY\_TAROT  	&\textBF{22.93}& \textBF{39.38} &-& -       			&\textBF{21.09}& \textBF{37.88} &-& -         &\textBF{20.51}&\textBF{37.58}&-& -   \\	
		\cmidrule{2-14}
		& CLUSTER'   	         &  \textBF{31.11}& \textBF{43.56} & \textBF{\textless 0.01}&0.12         &  \textBF{29.49}& \textBF{40.44 } & 0.16& 0.05       & 27.39& 36.84&  \textBF{\textless 0.01}& 0.10  \\
		& CLUSTER'\_TAROT &26.47& 42.98  &-& -         										&28.31&39.60&-& -       		 				 			&\textBF{28.96}&\textBF{41.26}&-& -\\
		
		\midrule
		
		\multirow{6}*{Infinispan} 
		& IR-ONLY  					&8.73& 25.44 & \textBF{\textless 0.01}& 0.09            &9.05& 26.84 & \textBF{\textless 0.01} & 0.10          		  & 7.32& 26.02& \textBF{\textless 0.01} & 0.14   \\
		& ConPOS              	   &9.45& 26.70  & \textBF{\textless 0.01}& 0.08          &9.48& 26.86  & \textBF{\textless 0.01}& 0.09            &7.68& 26.61 & \textBF{\textless 0.01}&0.13 \\
		& IR-ONLY\_TAROT  	 &\textBF{11.34}& \textBF{29.04}&-& -           			 &\textBF{12.00}& \textBF{30.68}&-& -          						&\textBF{10.17}&\textBF{27.82}&-& -   \\
		\cmidrule{2-14}
		& CLUSTER'   			  &  18.77& 30.50  & 0.13 & 0.04           						 &  \textBF{20.70}&  32.64  &0.35 &  0.02             			&20.92&   \textBF{31.82}&\textbf{0.02} &  0.06  \\
		& CLUSTER'\_TAROT   &\textBF{19.73}&\textBF{32.61}&-& - 			 			&20.69&\textBF{34.37}&-& -           	    						&\textBF{21.34}&31.12&-& - \\
			
		\midrule
			
		\multirow{6}*{Seam2} 
		& IR-ONLY  					&18.99& 40.61& \textBF{\textless 0.01}& 0.12             &17.19& 42.49& \textBF{\textless 0.01} & 0.10  		 &16.64&40.47&\textBF{\textless 0.01}& 0.11        \\
		& ConPOS                   &20.26& 41.96  &\textBF{\textless 0.01}& 0.11           &18.65& 43.49  & \textBF{\textless 0.01}& 0.10            &17.33& 42.01& \textBF{\textless 0.01}&0.11\\
		& IR-ONLY\_TAROT  &\textBF{23.65}&\textBF{46.11} &-& -                      &\textBF{21.85}&\textBF{46.41}&-& -           			 &\textBF{20.84}&\textBF{43.06}&-& -     \\
		\cmidrule{2-14}
		& CLUSTER'  			  &  34.16&  49.32  & 0.08 &  0.07           					&  33.45&47.47& \textBF{ 0.02} &  0.09          & 30.32& 45.95 & \textBF{0.02} &  0.09  \\
		& CLUSTER'\_TAROT   &\textBF{36.14}& \textBF{52.71}  &-& -                    &\textBF{37.06}&\textBF{52.54}&-& -  		   	   &\textBF{34.6}&\textBF{51.43}&-& -   \\
		
		\midrule
		
		\multirow{5}*{Drools} 
		& IR-ONLY  					&\textBF{8.87}&21.06& 0.12& 0.04           				&\textBF{7.98}& 20.98 & \textBF{0.02} & 0.07 				   &\textBF{7.56}&21.20& 0.14& 0.04     \\
		& ConPOS               	   &8.48& 21.27 & 0.13& 0.04          				       			&7.03& 21.41  & \textBF{0.02}& 0.07            					&6.67& 21.25&0.10&0.05 \\
		& IR-ONLY\_TAROT  	 &8.83& \textBF{22.41}  &-& -          			    	 &7.84& \textBF{22.46}  &-& -          									&7.39&\textBF{21.86} &-& -      \\
		\cmidrule{2-14}	
		& CLUSTER'   			  &  \textBF{15.86}&24.65 &\textBF{0.02}  &0.07         &13.62&  24.46  &\textBF{\textless 0.01} &  0.07    & \textBF{15.08}&  23.88 &  0.63 &  0.01  \\
		& CLUSTER'\_TAROT   &15.77&\textBF{26.72}   &-& -                   			  &\textBF{14.60}&\textBF{26.24}  &-& -         	           	&13.08& \textBF{24.24} &-& -   \\
			
		\midrule
		
		\multirow{6}*{Derby} 
		& IR-ONLY  				   &10.23&26.95& \textBF{\textless 0.01}& 0.20            	&9.58&26.74 & \textBF{\textless 0.01} & 0.17            &9.32& 26.21& \textBF{\textless 0.01} & 0.18           \\
		& ConPOS                  &11.58& 30.32  & \textBF{\textless 0.01} & 0.13           &10.31& 29.76  & \textBF{\textless 0.01}& 0.12            &8.38& 27.69& \textBF{\textless 0.01}&0.20 \\
		& IR-ONLY\_TAROT  	&\textBF{15.05}& \textBF{35.07} &-& -          				&\textBF{13.29}&\textBF{35.29} &-& -           &\textBF{14.22}&\textBF{34.01} &-& -     \\
		\cmidrule{2-14}
		& CLUSTER'               &24.81& 36.66&\textBF{\textless 0.01}&  0.06           & 24.11&  34.35  & \textBF{\textless 0.01} & 0.16           &23.67&  32.71&  \textBF{\textless 0.01}& 0.13  \\
		& CLUSTER'\_TAROT  &\textBF{25.05}&\textBF{43.51} &-& - 	                 &\textBF{27.91}&\textBF{42.23}   &-& - 			 							&\textBF{26.23}& \textBF{41.42} &-& - \\
		
		\midrule
		
		\multirow{6}*{Groovy} 
		& IR-ONLY  				   &26.98 & 52.72& \textBF{0.02}& 0.15         		    	&27.40&55.50 & \textBF{0.02} & 0.14         		 &19.69& 44.02& \textBF{\textless 0.01} & 0.22    \\
		& ConPOS                  &28.19& 56.39  & 0.47 & 0.04           				       &27.83& 58.99  & 0.52& 0.04            							&17.23& 48.61&0.89&0.01 \\
		& IR-ONLY\_TAROT  	&\textBF{33.59}& \textBF{58.96} &-& -                     &\textBF{33.53}&\textBF{61.80} &-& -                    &\textBF{27.91}&\textBF{54.44}&-& -    \\
		\cmidrule{2-14}
		& CLUSTER'   			 & 41.12&  69.75 &  \textBF{\textless 0.01} & 0.27      &  43.93&  73.69&\textBF{\textless 0.01}  &0.30        & 36.86&  64.82 &  \textBF{\textless 0.01} &  0.34  \\
		& CLUSTER'\_TAROT  & \textBF{46.49}&  \textBF{70.83}  &-& - 	             & \textBF{48.52} & \textBF{73.74}  &-& - 		            & \textBF{46.70}& \textBF{69.53 }&-& - \\
		
		\bottomrule
	\end{tabular}}
\end{table*}

\vspace{5pt}
To study RQ2, we choose \textbf{CLUSTER’} as the representative enhancing strategies built upon different perspectives for IR-based traceability recovery to test whether TAROT can collaborate with other enhancements.
CLUSTER’ propagates a small amount of user feedback with closeness analysis on code dependencies \cite{DBLP:journals/ese/GaoKMHLME22}.
We use IR-ONLY\_TAROT to generate IR candidate links and make them as the input of CLUSTER’. 
We name this combination of TAROT and CLUSTER’ as \textbf{CLUSTER’\_TAROT}.
Like RQ1, we also compared CLUSTER’ and  CLUSTER’\_TAROT on VSM, LSI, and JS, 
use F-measure at each recall point to see whether CLUSTER’\_TAROT can improve both accuracy and recall, 
use the Wilcoxon rank sum test to test $H_0$ ($\alpha$ = 0.05),
and use Cliff's $delta$ to quantify the amount of difference between CLUSTER’ and  CLUSTER’\_TAROT.

\begin{figure*}[t] 
\centering
\vspace{1pt}

\subfigure[iTrust-VSM]{
	\label{fig5.1.1}
	\includegraphics[width=0.135\textwidth]{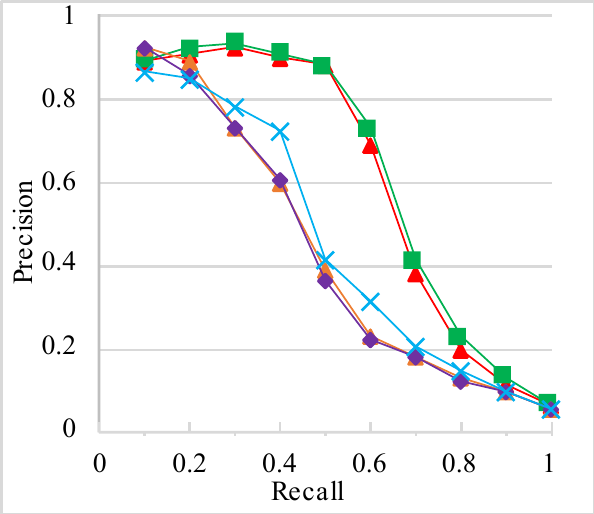}}
\subfigure[Gantt-VSM]{
	\label{fig5.1.2}
	\includegraphics[width=0.135\textwidth]{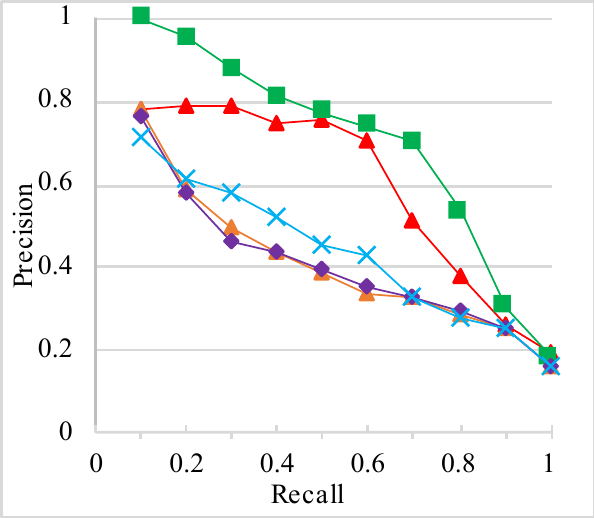}}
\subfigure[Maven-VSM]{
	\label{fig5.1.3}
	\includegraphics[width=0.133\textwidth]{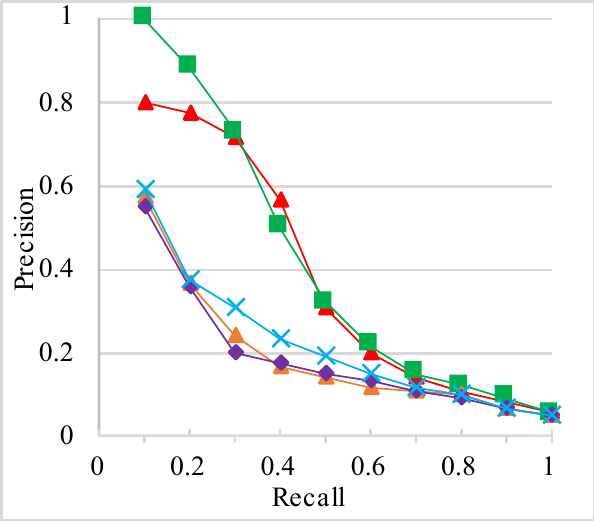}}
\subfigure[Pig-VSM]{
	\label{fig5.1.4}
	\includegraphics[width=0.13\textwidth]{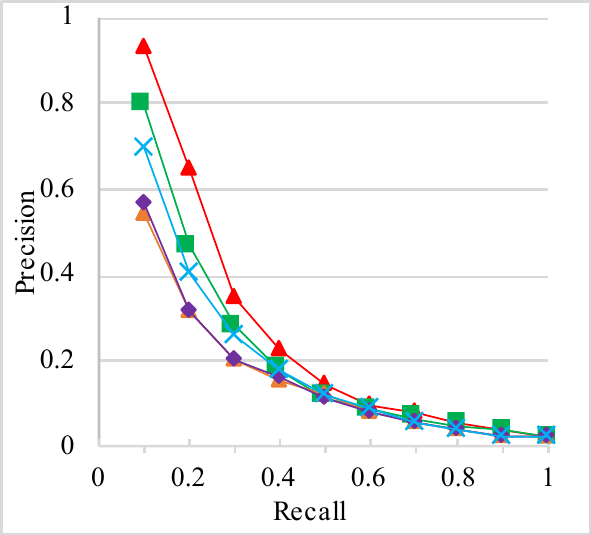}}
\subfigure[Infinispan-VSM]{
	\label{fig5.1.5}
	\includegraphics[width=0.13\textwidth]{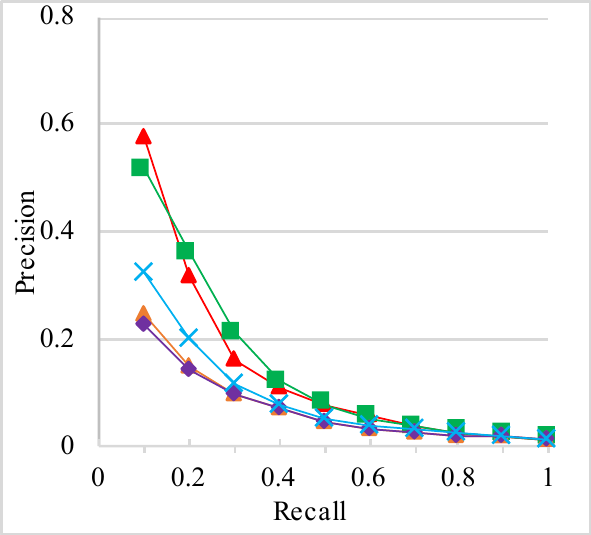}}
\subfigure[Seam2-VSM]{
	\label{fig5.1.6}
	\includegraphics[width=0.13\textwidth]{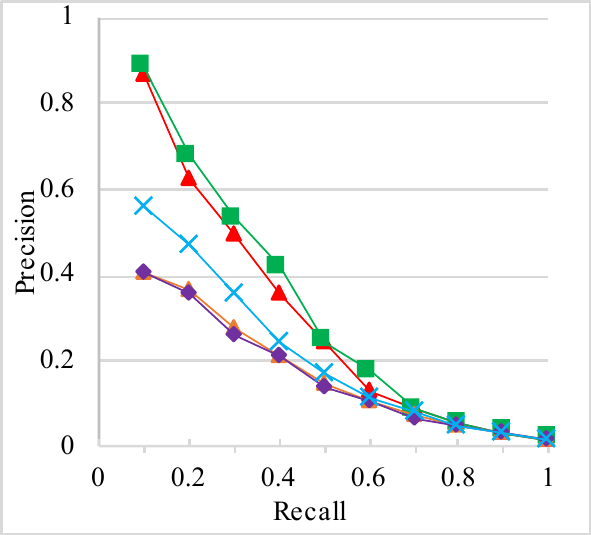}} 
\subfigure[Drools-VSM]{
	\label{fig5.1.7}
	\includegraphics[width=0.13\textwidth]{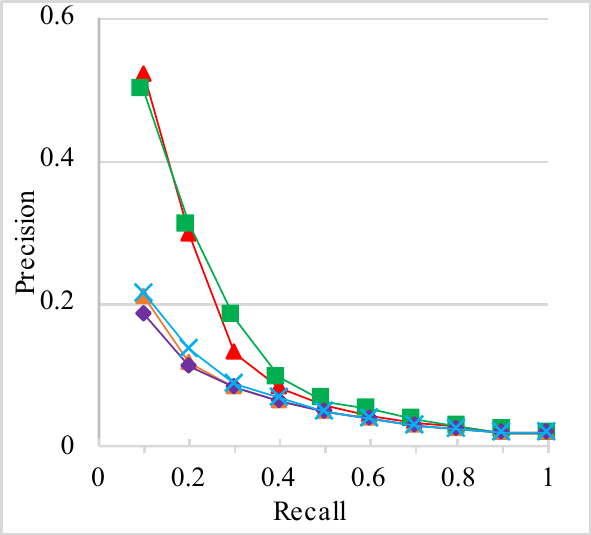}}\\

\subfigure[iTrust-LSI]{
	\label{fig5.1.8}
	\includegraphics[width=0.135\textwidth]{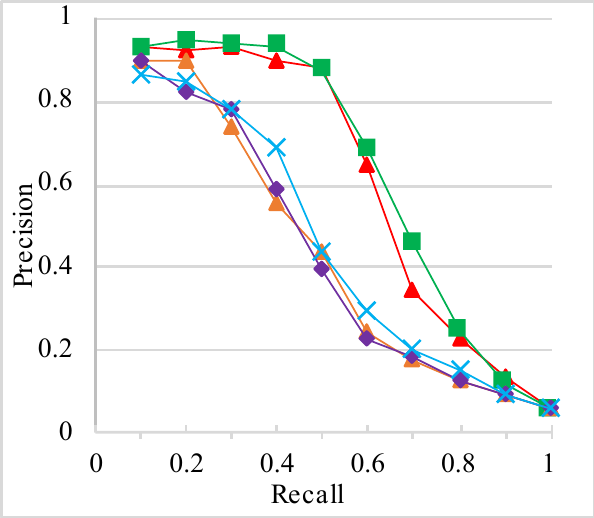}}
\subfigure[Gantt-LSI]{
	\label{fig5.1.9}
	\includegraphics[width=0.135\textwidth]{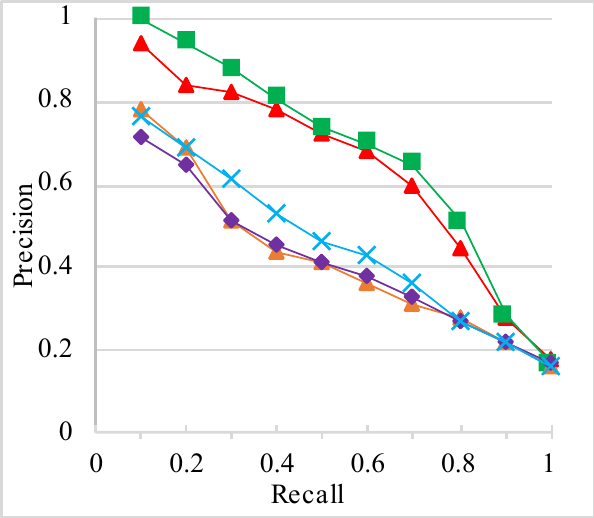}}
\subfigure[Maven-LSI]{
	\label{fig5.1.10}
	\includegraphics[width=0.133\textwidth]{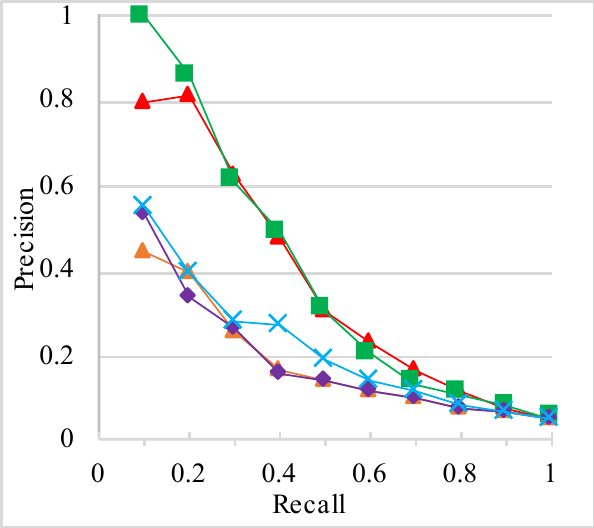}}
\subfigure[Pig-LSI]{
	\label{fig5.1.11}
	\includegraphics[width=0.13\textwidth]{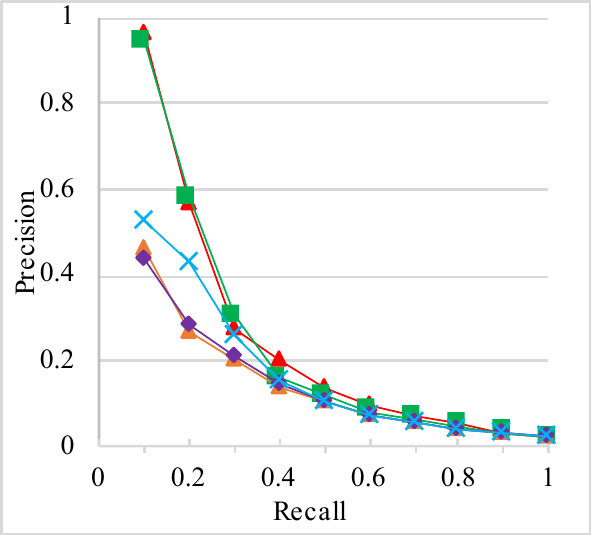}}
\subfigure[Infinispan-LSI]{
	\label{fig5.1.12}
	\includegraphics[width=0.13\textwidth]{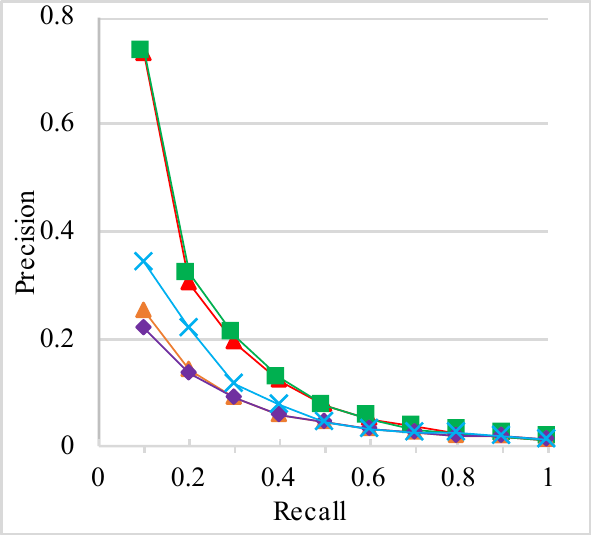}}
\subfigure[Seam2-LSI]{
	\label{fig5.1.13}
	\includegraphics[width=0.13\textwidth]{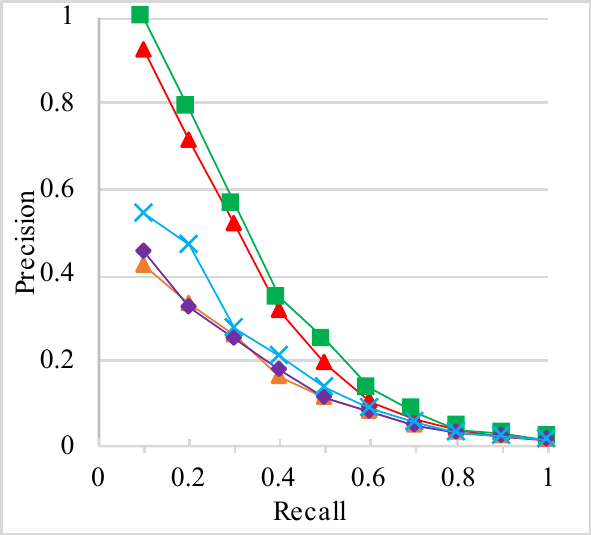}}
\subfigure[Drools-LSI]{
	\label{fig5.1.14}
	\includegraphics[width=0.13\textwidth]{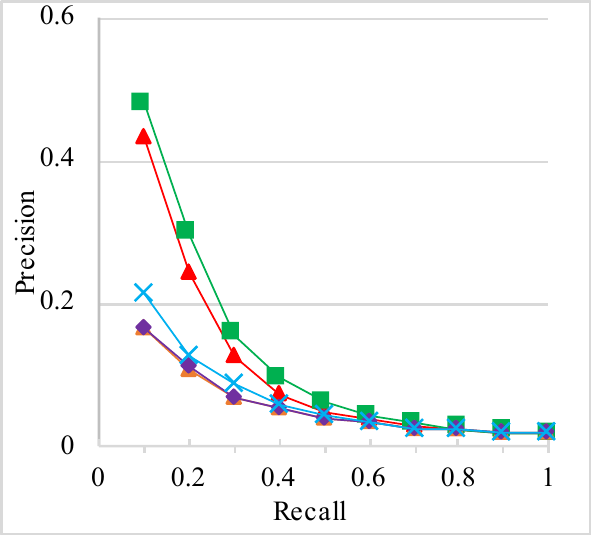}}\\

\subfigure[iTrust-JS]{
	\label{fig5.1.15}
	\includegraphics[width=0.135\textwidth]{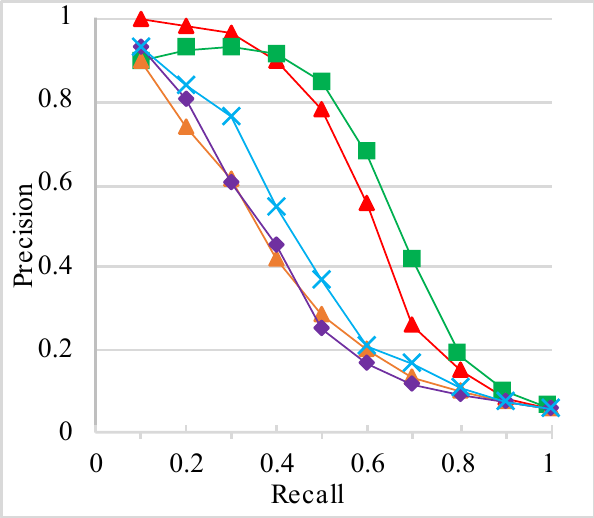}} 
\subfigure[Gantt-JS]{
	\label{fig5.1.16}
	\includegraphics[width=0.135\textwidth]{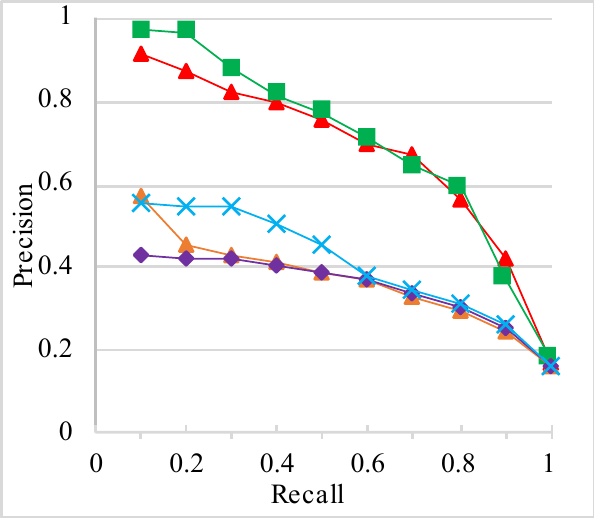}}
\subfigure[Maven-JS]{
	\label{fig5.1.17}
	\includegraphics[width=0.133\textwidth]{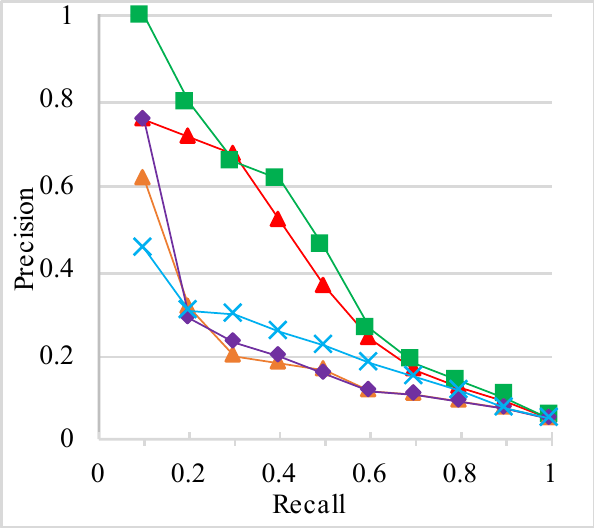}}
\subfigure[Pig-JS]{
	\label{fig5.1.18}
	\includegraphics[width=0.13\textwidth]{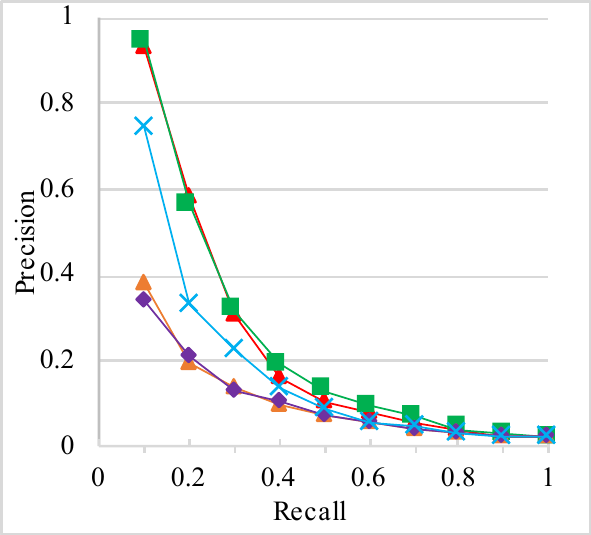}}
\subfigure[Infinispan-JS]{
	\label{fig5.1.19}
	\includegraphics[width=0.13\textwidth]{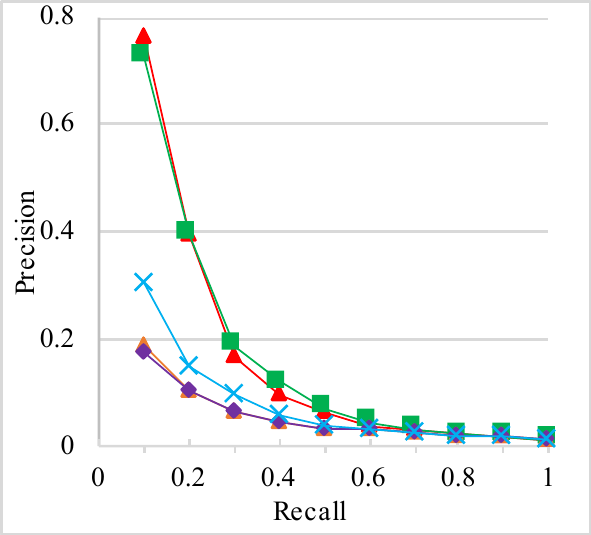}}
\subfigure[Seam2-JS]{
	\label{fig5.1.20}
	\includegraphics[width=0.13\textwidth]{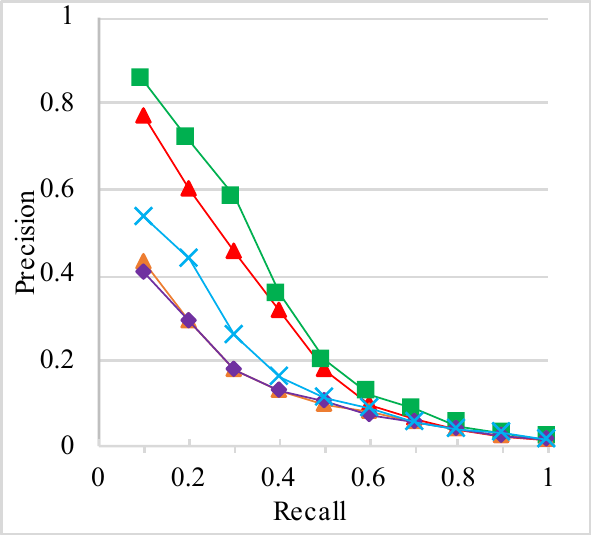}}
\subfigure[Drools-JS]{
	\label{fig5.1.21}
	\includegraphics[width=0.13\textwidth]{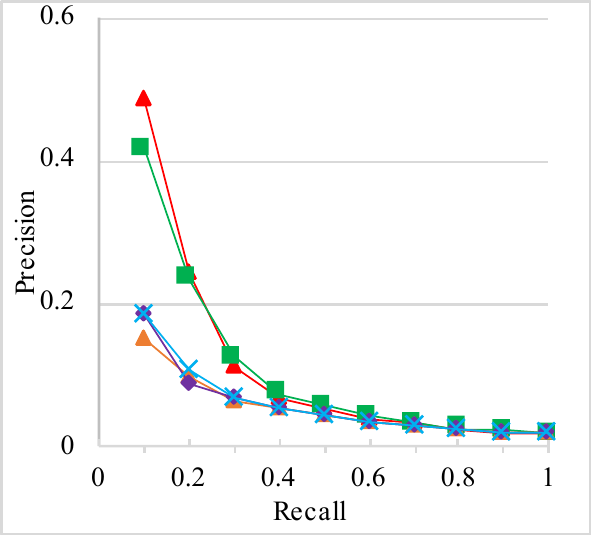}} \\

\subfigure[Groovy-VSM]{
	\label{fig5.1.22}
	\includegraphics[width=0.135\textwidth]{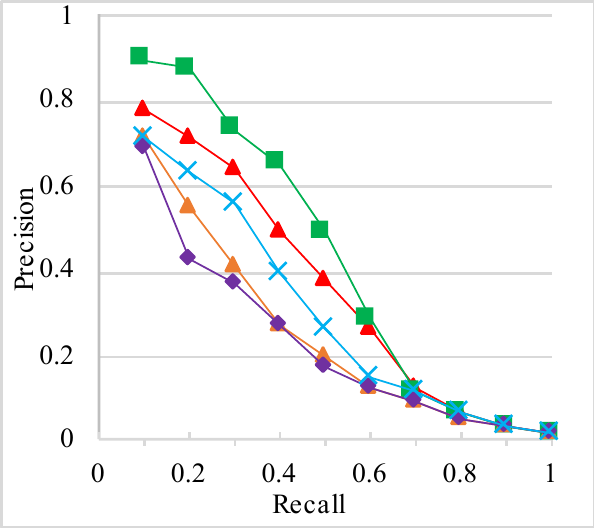}}
\subfigure[Groovy-LSi]{
	\label{fig5.1.23}
	\includegraphics[width=0.135\textwidth]{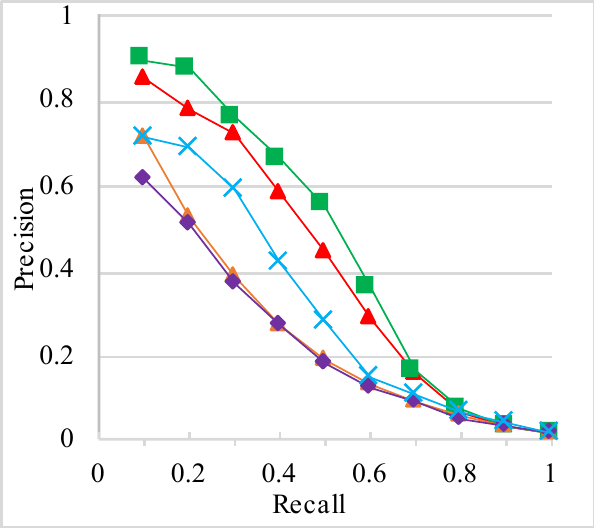}}
\subfigure[Groovy-JS]{
	\label{fig5.1.24}
	\includegraphics[width=0.135\textwidth]{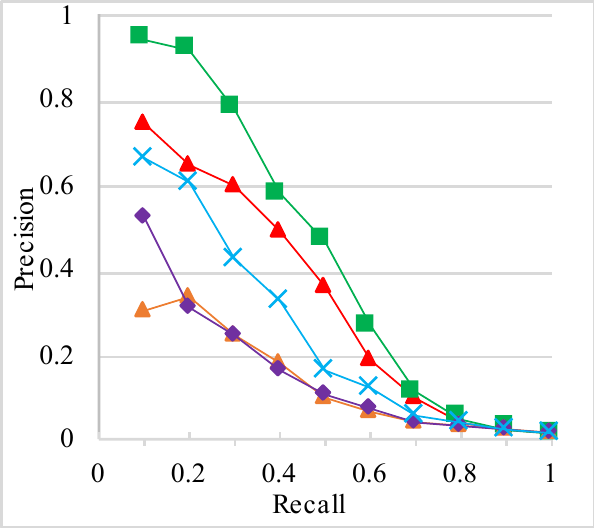}}
\subfigure[Derby-VSM]{
	\label{fig5.1.25}
	\includegraphics[width=0.132\textwidth]{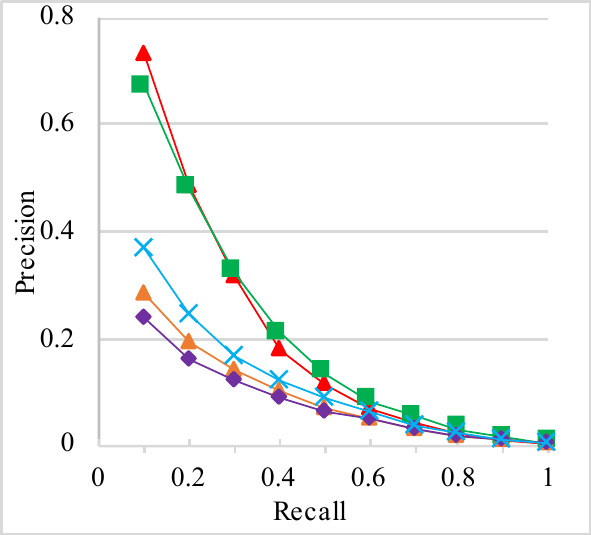}}
\subfigure[Derby-LSI]{
	\label{fig5.1.26}
	\includegraphics[width=0.132\textwidth]{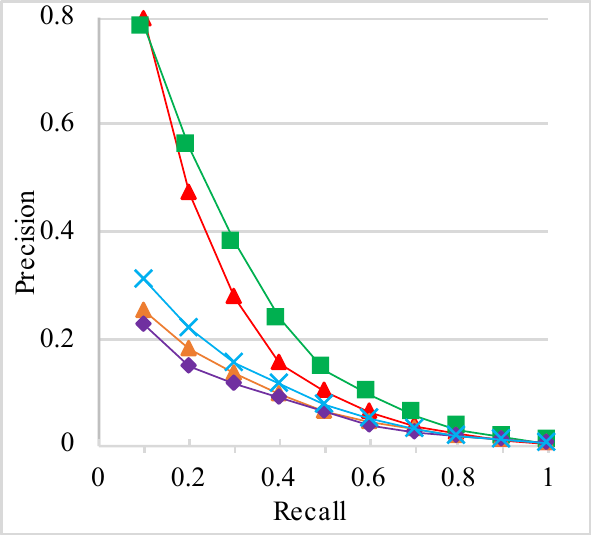}} 
\subfigure[Derby-JS]{
	\label{fig5.1.27}
	\includegraphics[width=0.132\textwidth]{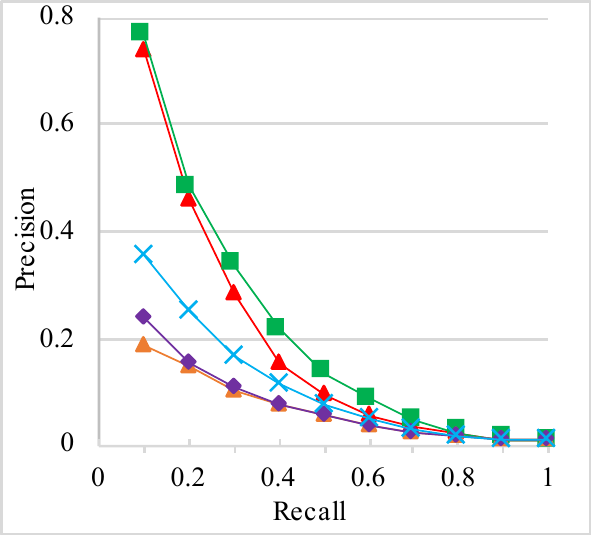}}\\

\includegraphics[width=0.75\textwidth]{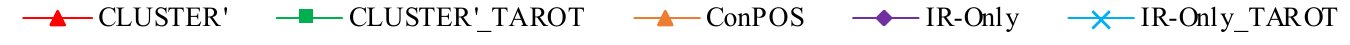} \\

\caption{Precision/Recall curves grouped by evaluated systems (subgraph (a), (h), (o) for iTrust, and similarly for Gantt, Maven, Pig, Infinispan, Seam2 and Drools, while (v), (w), (x) for Groovy, and similarly for Derby) and IR models (VSM, LSI, and JS).}
\label{fig5.1}
\end{figure*}

\vspace{5pt}
To study RQ3,  we perform an ablation study to evaluate the performance of TAROT's different parts.
We regard IR-ONLY as the basics and add each part of TAROT to the basics incrementally.
By comparing the performance of the approach before and after adding a part can see how many contributions that each step of TAROT makes.
Specifically, “$+ b$” represents when we only use extracted consensual biterms to improve the input quality for IR techniques, without considering the \textit{global weight} $\lambda$ and the \textit{local weight} $\theta$.
“+ $\lambda$” represents when we adjust IR values through the biterms, the \textit{global weight} $\lambda$ will be taken into consideration.
“+ $\theta$” represents when we adjust IR values through the biterms, the \textit{local weight} $\theta$ will be taken into consideration.
We also use VSM, LSI, and JS, to compare IR-ONLYs with different steps included.


\section{Discussions on results}

\textbf{RQ1: Can TAROT help improve the performance of IR-based traceability recovery?}
Table \ref{tab5.1} shows the experiment results of nine evaluated systems (rows).
For each system and each IR model (columns), we compared the performance of  IR-ONLY, ConPOS, and IR-ONLY$\_$TAROT.
Sub column 1 shows the average precision (AP).
Sub column 2 shows the mean average precision (MAP).
Sub column 3 shows the $p$-value of the F-measure significance test for IR-ONLY$\_$TAROT and sub column 4 shows the Cliff’s $\delta$.
For each approach, the best result of AP, MAP, and $p$-value $\leq$ 0.05 are in bold text.
The results show that IR-ONLY$\_$TAROT outperforms ConPOS on both AP and MAP for all cases.
When compared with IR-ONLY, IR-ONLY$\_$TAROT performs better on MAP for all case, only performs worse on AP in 3 cases of DroolsAP on Drools (0.12 on average).
Specifically, in 42 out of 54 cases, the F-measure for the result of IR-ONLY$\_$TAROT is significantly higher than those of the two baseline approaches (p-value $\leq$ 0.05) at each level of recall, which indicates that IR-ONLY$\_$TAROT outperforms baseline approaches in the majority of cases.
Figure \ref{fig5.1} shows and compares the precision-recall curves for the three approaches grouped by each system and IR model.

We now use the adapted excerpt from the iTrust system to demonstrate why IR-ONLY$\_$TAROT is able to outperform IR-ONLY and ConPOS.
For use case UC35 and code \texttt{EmailUtil} as shown in Figure \ref{fig1.1}, they shared two biterms, i.e., “\textbf{sendEmail}” and “\textbf{fakeEmail}”. 
Because the two biterms both appears in sub-flows of UC35 twice, TAROT adds them into UC35 twice respectively according to the biterms extension rules (discussed in Section \ref{sec3.2.1}).
For \texttt{EmailUtil}, TAROT adds “\textbf{sendEmail}” three times (appears in class comment and method name) and “\textbf{fakeEmail}” once (appears in invoked method name). 
It turns out that TAROT can successfully enrich texts of UC35 and \texttt{EmailUtil}.
Furthermore, TAROT further uses \textit{global weight} $\lambda$ and \textit{local weight} $\theta$ to improve the IR value of “UC35 -> \texttt{EmailUtil}”. 
For the result of IR-ONLY, the ranks of \texttt{EmailUtil} in the candidate list of UC35 and lists of all 34 use cases are 40/137 and 1172/4658 respectively, while IR-ONLY$\_$TAROT promotes them to 8/137 and 201/4658 respectively.
Yet for ConPOS, there is only one verb shared by UC35 and \texttt{EmailUtil}, i.e., “send", which does not help to achieve much improvement.

\begin{table}[tb]
	\caption{The number of computed AP and MAP evaluating each part of TAROT ($b$ = consensual biterms, $\lambda$ = gloabal weight, and $\theta$ = local weight)}
	\renewcommand\arraystretch{0.6}
	\resizebox{\linewidth}{!}{
	\tiny
	\label{tab5.2}
	\begin{tabular} {lcllllll}
		\toprule
		&         & \multicolumn{2}{c}{VSM} & \multicolumn{2}{c}{LSI} & \multicolumn{2}{c}{JS} \\ 
		\cmidrule(r){3-4} 	\cmidrule(r){5-6}	\cmidrule(r){7-8}
		&          &AP& MAP       &AP& MAP        &AP& MAP      \\ 
		\midrule
		
		\multirow{5}*{iTrust}  
		& IR-ONLY		        	     &45.78& 58.43       		           &46.01& 59.17           		        &40.57& 56.01  \\
		& + $b$ 							 &48.28& 58.43   			           &47.85& 59.26          				&44.66&58.40 \\
		& + $b$ + $\lambda$  	   	         &49.16& 60.49         		           &49.04& 59.72       	                &\textBF{45.89}&\textBF{58.71 } \\	
		& + $b$ + $\lambda$  +$\theta$     &\textBF{49.50}& \textBF{62.12}       &\textBF{49.18}& \textBF{61.50}      &45.74&58.59\\
		\midrule
	
		\multirow{5}*{GanttProject}    
		& IR-ONLY      					&43.17& 49.79  						   &43.89	& 51.72   					&36.50& 46.76    \\
		& + $b$     						&45.72& 53.28  						   &46.19	& 53.24 					&41.23& 50.81           \\	
		& + $b$ + $\lambda$  	   			&46.19& 53.26         				   &46.92& 54.43     					&41.38&50.76 \\		
		& + $b$ + $\lambda$  +$\theta$    &\textBF{47.37}& \textBF{54.09}        &\textBF{48.63}& \textBF{55.25}      &\textBF{43.90}&\textBF{51.76}  \\
		
		\midrule
		
		\multirow{5}*{Maven} 
		& IR-ONLY  				    	&22.27& 37.11         				   &22.15&40.65              			&24.08& 40.29        \\
		& + $b$  							&24.42& 41.47           			   &22.75&45.27            				&25.81& 43.39          \\
		& + $b$ + $\lambda$   	   		&26.64& \textBF{44.22}   			   &25.60& 46.81        				&\textBF{26.01}&47.05 \\
		& + $b$ + $\lambda$  +$\theta$    &\textBF{26.77}&43.65            	   &\textBF{26.97}& \textBF{47.17}      &25.50&\textBF{48.35}\\

		\midrule
		
		\multirow{5}*{Pig} 
		& IR-ONLY    	           		&19.71& 36.37          				   &17.89& 36.62           			 	&14.64& 31.90      \\
		& + $b$    	    				&21.55& 37.75        				   &19.52& 36.92        		 		&17.70& 34.35       \\
		& + $b$ + $\lambda$   	   		&22.23& 37.61        				   &20.41& 36.60          				&18.44&34.98   \\
		& + $b$ + $\lambda$  +$\theta$    &\textBF{22.93}& \textBF{39.38}        &\textBF{21.09}& \textBF{37.88}      &\textBF{20.51}&\textBF{37.58}  \\

		\midrule
			
		\multirow{5}*{Infinispan} 
		& IR-ONLY  						&8.73& 25.44            			   &9.05& 26.84           		  		& 7.32& 26.02   \\
		& + $b$  							&9.76& 27.53            			   &9.99& 28.72         				& 8.25& 27.81        \\	
		& + $b$ + $\lambda$  	   			&10.71& 28.29     					   &11.05& 29.75     					&9.40&27.66  \\	
		& + $b$ + $\lambda$  +$\theta$    &\textBF{11.34}& \textBF{29.04}        &\textBF{12.00}& \textBF{30.68 }     &\textBF{10.17}&\textBF{27.82}  \\
		
		\midrule
			
		\multirow{5}*{Seam2} 
		& IR-ONLY  						&18.99& 40.61           			   &17.19& 42.49 		 				&16.64&40.47       \\
		& + $b$  							&21.58& 43.25                  		   &17.81& 42.70		 				&17.71&39.95        \\
	    & + $b$ + $\lambda$   	  		&\textBF{24.05}&44.76         		   &21.34&45.14        					&20.53 &42.26   \\		
		& + $b$ + $\lambda$  +$\theta$    &23.65&\textBF{46.11 }          	   &\textBF{21.85}&\textBF{46.41}       &\textBF{20.84}&\textBF{43.06} \\
		
		\midrule
		
		\multirow{5}*{Drools} 
		& IR-ONLY  						&8.87&21.06          				   &7.98& 20.98 				   		&7.56&21.20    \\
		& + $b$							&\textBF{9.12}&\textBF{22.57}          &\textBF{8.75}& 23.10    		    &\textBF{7.68}&21.43          \\		
		& + $b$ + $\lambda$   	  	    &8.99& 22.10       					   &8.45& \textBF{23.37}       			&7.47&21.54  \\	
		& + $b$ + $\lambda$  +$\theta$    &8.83& 22.41         				   &7.84& 22.46       				    &7.39&\textBF{21.86}   \\
		
		\midrule
			
		\multirow{5}*{Derby} 
		& IR-ONLY  				  		&10.23&26.95          				   &9.58&26.74           				&9.32& 26.21          \\
		& + $b$  							&12.42&30.83            			   &10.18&28.88          				&10.63& 28.50          \\
		& + $b$ + $\lambda$  	   			&13.78& 33.21       				   &11.77&32.64         				&12.26&31.71 \\		
		& + $b$ + $\lambda$  +$\theta$    &\textBF{15.05}& \textBF{35.07 }       &\textBF{13.29}&\textBF{35.29}       &\textBF{14.22}&\textBF{34.01} \\
		\midrule
		
		\multirow{5}*{Groovy} 
		& IR-ONLY  				   		&26.98 & 52.72         		    	   &27.40&55.50        		 			&19.69& 44.02\\
		& + $b$  							&27.94 & 52.27     					   &28.08&54.69           				&22.15& 46.52       \\		
		& + $b$ + $\lambda$  	   			&31.34& 55.99    					   &31.75&58.11          				&26.93&52.94 \\	
		& + $b$ + $\lambda$  +$\theta$     &\textBF{33.59}& \textBF{58.96}       &\textBF{33.53}&\textBF{61.80} 		&\textBF{27.91}&\textBF{54.44} \\
	
		\bottomrule
	\end{tabular}}

\end{table}

\textbf{RQ2: Can TAROT collaborate with other enhancing strategies for IR-based traceability recovery? }Table \ref{tab5.1} shows the performances of CLUSTER’ and CLUSTER’ $\_$TAROT on nine evaluated systems.
In 15 out of 27 cases, the F-measure for the result of CLUSTER’ $\_$TAROT is significantly higher than that of CLUSTER’ (p-value $\leq$ 0.05) at each level of recall, which indicates that CLUSTER’ $\_$TAROT outperforms CLUSTER’ in the majority of cases.
When compared with CLUSTER’, CLUSTER’ $\_$TAROT outperforms in 22 out of 27 cases on AP.
Its AP is 2.49 higher than that of CLUSTER’ on average in total 27 cases.
The highest value is 10.69 on GanttProject-VSM.
Meanwhile, CLUSTER’ $\_$TAROT outperforms CLUSTER’ in 23 out of 27 cases on MAP.
Its MAP is 3.16 higher than that of CLUSTER’ on average in total 27 cases.
The highest value is 8.71 on Derby-JS.
It turns out that TAROT can collaborate with CLUSTER', which is a totally different IR enhancing strategies built upon both code dependency analysis and the use of user feedback. The combined  CLUSTER’\_TAROT can not only achieve further improvement where CLUSTER' has already achieves good results (e.g., iTrust and GanttProject), but also achieve indeed improvement where CLUSTER' achieves relatively low results (e.g., Infinispan and Drools). 
Finding more effective way to collaborate TAROT with other enhancing strategies is one of our future work.
This result also inspires us to further explore whether TAROT can be beneficial to ML-based techniques in our future work, because TAROT inherently improves the quality and quantity of requirement and code texts.

\textbf{RQ3: How many contributions does different parts of TA-ROT make individually?}
Table \ref{tab5.2} shows the experiment results of nine evaluated systems (rows).
For each system and each IR model (columns), we compared the performance of incrementally adding each part of TAROT including “+ $b$" means only extending artifact texts with consensual biterms on the bases of the pure IR-based approach (IR-ONLY), “+ $b$ + $\lambda$" means further introducing \texttt{global weight} $\lambda$, and “+ $b$ + $\lambda$ + $\theta$" means the entire TAROT.
Sub column 1 shows the average precision (AP).
Sub column 2 shows the mean average precision (MAP).
For each approach, the best result of AP, MAP, and $p$-value $\leq$ 0.05 are in bold text.
From the table, we can observe that “$+b$" outperforms IR-ONLY in all 27 cases on AP (on average 1.69) and in 24 out of 27 cases on MAP (on average 2.18) , which indicates that TAROT can extract funcationality consensual biterms and promote the quality of artifact texts. 
In addition, the results show that “+ $b$ + $\lambda$" outperforms “$+b$" in 24 out of 27 cases on AP (on average 1.69) and 21 out of 27 cases on MAP (on average 2.02) respectively, which means shared consensual biterms can narrow down the semantic gap.
Moreover, the results show that “$b$ + $\lambda$ + $\theta$" can outperform “+ $b$ + $\lambda$" in 21 out of 27 cases on AP  (on average 1.13) and 24 out of 27 cases on MAP (on average 1.43) respectively.
We also noted that all three parts of TAROT can outperform IR-ONLY in almost all cases on both AP and MAP, only in 4 out of 162 cases performs worse.

The overall evaluation results also implied that TAROT is likely to be beneficial in practice because: (1) it can improve IR-based traceability by solely analyzing the text structures without requiring additional analysis (such as code instrumentation, which may not be viable for real-world projects);  (2) it can still work even on systems with really low-quality texts (e.g., Infinispan, Drools, and Derby in our evaluation); (3) it can complement other enhancing strategies, and is also likely to be useful for ML-based traceability recovery because TAROT inherently improves the quality and quantity of requirement and code texts.

\section{Threats to Validity}

\textbf{Internal threats. }One possible internal threat to our approach is that we cannot guarantee 100\% accuracy in segmenting texts, tagging POSs, and parsing dependencies based on Stanford CoreNLP. 
However, existing work has reported that the accuracy of off-the-shelf NLP tools is acceptable when analyzing texts with the context of proper sentences and grammatical structures \cite{DBLP:journals/infsof/AliCHH19}, and we found no obvious errors in the output of Stanford CoreNLP as well. We will consider SE-specific NLP tools (e.g., \citep{DBLP:conf/icse/PartachiDTB20}) in future work.

\textbf{External threats. }Our evaluation uses the same dataset of CLUSTER' \cite{DBLP:journals/ese/GaoKMHLME22} with nine systems including seven open-source systems.
We think that our findings from this evaluation are relevant because these evaluated systems are either widely studied or used in practice from different domains \cite{DBLP:conf/iwpc/KuangG0M0ME19, DBLP:conf/re/RathRM17}. 
Furthermore, we combined the evaluated systems with three mainstream IR models (i.e., VSM, LSI, and JS) to extend our experiments to a total of 27 variants (e.g., iTrust-VSM and Pig-JS). 
However, one possible threat is that our evaluation is only based on part of the system functionalities of Maven, Pig, Infinispan, Drools, Derby, Seam2, and Groovy because Gao et al. set up heuristics to filter and merge issues with linked commit logs from the IlmSeven dataset to elicit both requirements and RTMs, thus covering only part of the extracted issues from Jira and GitHub. Nevertheless, we use the same RTMs to compare our approach with the baseline approaches on each evaluated system, thus making no bias.

\section{Conclusions and Future Work}

In this study, we propose to extract co-occurred word pairs from the text structures of both requirements and code (i.e., consensual biterms) to improve IR-based traceability recovery. The consensual biterms are extracted by our crosscheck between candidate requirement biterms and candidate code biterms. We argue that although the consensual biterms are just correlative term combinations that can be located in both requirements and code, they represents important semantics of the system that are consistently followed during the software development process, thus being valuable for automated traceability recovery. We then use these biterms to first enrich the input corpora for IR techniques, and then adjust the generated IR values.
An empirical evaluation based on nine real-world systems shows that our approach can not only outperform baseline approaches, but also collaborate with other enhancing strategies built upon different perspectives.
In future work we first plan to extract more consensual biterms from additional software artifacts (e.g., design and tests). We then plan to study whether consensual biterms can improve ML-based traceability recovery as well, such as improving IR values used as features in ML-based approaches, or enhancing word-embeddings with additional biterms.
Our replication package is available at: https://github.com/huiAlex/TAROT.

\begin{acks}
Ye Zhang also contributes to this work. 
This work is supported by the National Natural Science Foundation of China (No.62072227, No.62025202), the National Key Research and Development Program of China (No.2019YFE0105500) jointly with the Research Council of Norway (No.309494), the Key Research and Development Program of Jiangsu Province (No.BE2021002-2), the Austrian Science Fund (FWF, grants P31989 and I4744), the German Ministry of Education and Research (BMBF) grants 01IS18074E, as well as the Intergovernmental Bilateral Innovation Project of Jiangsu Province (No.BZ2020017). Hongyu Kuang is the corresponding author. 

\end{acks}

\bibliographystyle{ACM-Reference-Format}
\bibliography{reference}

\end{document}